\documentclass[3p,review,12pt]{elsarticle}

\usepackage{amssymb}

\usepackage{lineno}

\usepackage{float}
\usepackage{color}
\usepackage{svg}
\usepackage{epstopdf}
\usepackage{amsfonts}
\usepackage{amsmath}
\usepackage{pdflscape}
\usepackage{graphicx}
\usepackage{pdfpages}
\usepackage{subcaption}
\usepackage{bbm,bm}
\usepackage{stmaryrd}
\usepackage{color}
\usepackage{booktabs}

\usepackage{yhmath}

\usepackage{flexisym}
\usepackage{adjustbox}
\usepackage{physics}
\usepackage{boldline,multirow}

\graphicspath{{./figures/}}
\usepackage{tabularx}
\usepackage{upgreek}

\usepackage[linesnumbered,ruled,vlined]{algorithm2e}
\usepackage[colorlinks,bookmarksopen,bookmarksnumbered,citecolor=blue,urlcolor=red]{hyperref}
\journal{Chemical Engineering Journal}

\usepackage{etoolbox}
\makeatletter
\patchcmd{\pprintMaketitle}
  {\unvbox\keybox\par\vskip10pt\fi\hrule}
  {\unvbox\keybox\par\vskip10pt\fi}
  {}{}

\patchcmd{\MaketitleBox}
  {\unvbox\keybox\par\vskip10pt\fi\hrule}
  {\unvbox\keybox\par\vskip10pt\fi}
  {}{}
\makeatother

\begin{document}

\begin{frontmatter}

\title{A hierarchical graph neural network for efficient particle-resolved prediction of granular flow dynamics}

\author[1]{Kuang Shi}
\author[2]{Qingqi Zhao}
\author[3]{Teng Man}
\author[4]{Sergio Andres Galindo-Torres}
\author[1]{Yimin Lu \texorpdfstring{\corref{cor}}{}}
\ead{Yimin.Lu@ttu.edu}

\cortext[cor]{Corresponding author}
\address[1]{Department of Civil, Environmental, and Construction Engineering, Texas Tech University, 79409, USA}
\address[2]{Stuttgart Center for Simulation Science, University of Stuttgart, Stuttgart, 70569, Germany}
\address[3]{College of Civil Engineering, Zhejiang University of Technology, Hangzhou, 310023, Zhejiang, China}
\address[4]{Key Laboratory of Coastal Environment and Resources of Zhejiang Province, School of Engineering, Westlake University, Hangzhou, 310024, Zhejiang, China}

\begin{abstract}
Graph neural surrogates for granular flow can resolve the motion of individual grains, but propagating information across a large amount of grains requires repeated, costly updates on the full particle graph. This study introduce a Hierarchical Coarse-Grained Graph Neural Network (HCG-GNN) that couples direct interactions between neighboring grains with information exchange on two coarser graphs of occupied spatial cells. In discrete element method (DEM) benchmarks of granular flows down incline planes, HCG-GNN captured key flow characteristics, including front propagation, lateral spreading, deposition morphology, free surface evolution, internal shear behavior, and energy conversion. Results also show that preserving grain-specific features as coarse information returned to the particle graph and integrating predicted accelerations to advance particle motion both improved multi-step accuracy. With 5 of its 7 message-passing steps performed on coarse graphs, HCG-GNN achieved accuracy comparable to a conventional particle-level GNN with a 6.47-fold training speedup and 46.2\% lower GPU memory use. A model trained on up to 5000 grains predicted a 2~s release of 10000 grains in 31~s, compared with 2000~s for DEM under the benchmark conditions. This combination of speed and grain-scale resolution supports particle-resolved inverse design, with prescribed runout or discharge targets guiding the search for suitable geometries and grain dynamics revealing the basis for differences in performance.
\end{abstract}


\end{frontmatter}

\section{Introduction}
\label{sec_intro}

Granular flows occur in a wide range of natural hazards and industrial processes, including landslides and debris flows~\cite{zarch2025rheological,lu2026hydro}, transport and handling of bulk solids \cite{deak2025high}, and additive manufacturing~\cite{escano2018revealing,nan2020numerical}. Across these settings, macroscopic quantities such as flow depth, discharge rate, and runout distance emerge from the collective motion of grains as contacts form, persist, and break. Laboratory experiments provide direct observations of granular flow and deposition, although access to the interior of a dense assembly is often limited \cite{li2026generative}. Continuum mechanics models offer an efficient description of bulk motion, but their accuracy depends on whether the constitutive law captures the material response across the stresses and shear rates encountered during flow and jamming \cite{wang2025stateboundary,jin2022fidelity,lu2024shear}. The discrete element method (DEM) resolves individual particle trajectories and their collective response by explicitly computing particle motion and contact forces. However, this resolution comes at substantial computational cost since contact interactions need to be updated throughout the assembly at time increments small enough to resolve their dynamics. This makes particle-scale modeling particularly demanding in parameter studies, design optimization, and inverse analysis, where many forward simulations are required~\cite{jiang2024inverse,choi2024inverse}.

Data-driven methods have been applied to a broad range of granular material problems~\cite{qu2025ai}, including porosity prediction~\cite{liu2020porosity,bui2019demann}, packing density optimization~\cite{baule2023packing}, force-chain prediction~\cite{cheng2024forcechains}, and reconstruction of pressure and velocity fields~\cite{hu2026pressure,li2026generative}. Applications also extend to the discovery of governing equations~\cite{zhao2023governing}, identification of rheological parameters~\cite{baldoni2026rheological}, constitutive modeling~\cite{han2026finn,irani2026ginnf,bahmani2024elastoplasticity,vlassis2022component}, and prediction or control of particle-handling processes such as screening and manipulation~\cite{sun2027length,wei2026manipulation}. However, distinct grain arrangements can yield similar bulk quantities or spatially averaged fields, so surrogates based solely on these bulk-level descriptors cannot distinguish how individual grains are connected to their neighbors. Graph neural networks (GNNs) can accommodate this evolving connectivity by representing grains as nodes and neighboring pairs as edges, along which information is exchanged to predict particle accelerations~\cite{sanchez2020learning,choi2024gnn}. Particle-based GNNs have been adapted to complex geometries and moving boundaries~\cite{li2024advanced,le2026dynamic}, extended through temporal and equivariant formulations~\cite{zhao2025temporal,huang2026equivariant}, and applied to granular mixing and high-energy ball milling~\cite{yang2026bayesian,garrido2026ballmilling}. However, in a conventional particle graph, each round of message passing extends the range of information exchange by only one edge. Propagating information over larger physical distances thus requires either additional passes over the full particle graph or a larger neighborhood for each grain, both of which increase computational and memory demands~\cite{fransen2025sciml}.

Multiscale and reduced-order methods in computational mechanics offer several ways to reduce the cost of a fine-scale representation \cite{karapiperis2021multiscale,storm2024microstructure}. In granular systems, coarse-grained DEM reduces the number of simulated particles by introducing larger representative grains~\cite{zhang2024microscopic}, while subspace graph networks project particle motion onto a lower-dimensional basis for learning~\cite{haeri2024subspace}. Graph coarsening offers a further possibility: individual particles remain the units of prediction, while part of the computation takes place on a smaller set of nodes that aggregate information over larger spatial regions. Hierarchies of this kind have been used to represent interactions in particle systems~\cite{martinkus2021scalable}, exchange information among material points through sparse grids~\cite{chentanez2022learning}, and accelerate mesh-based simulations~\cite{perera2024multiscale}. For granular flow, the difficulty is to retain information about the spatial arrangement of grains~\cite{vlassis2024sand} as the occupied region changes and deformation localizes within the assembly~\cite{tordesillas2022localization}. For example, a coarse cell spanning a shear layer may contain grains with similar positions but markedly different velocities. Their mean motion can describe transport through the cell while obscuring the velocity gradient within it. Therefore, the accuracy retained under coarsening must be assessed at both scales, through the evolution of the bulk flow and the variation of particle motion within the flowing layer.

To address these challenges, we develop a Hierarchical Coarse-Grained Graph Neural Network (HCG-GNN) for particle-resolved simulation of granular flow. The architecture retains direct message passing between neighboring grains before pooling and after coarse information is returned to the particles. Between these two grain-level updates, information propagates through two successively coarser graphs constructed from occupied spatial cells. Each particle acceleration is then predicted using both its local neighborhood and information from a larger region of the assembly. The model is evaluated against DEM simulations of granular releases down an incline onto a horizontal bed with different particle numbers, inclination angles, and initial packing conditions. Its performance is assessed through autoregressive rollout accuracy and comparisons of front propagation, particle redistribution, surface profiles, internal deformation, and energy evolution. Comparisons with a conventional particle-level GNN and variations in hierarchy depth and resolution quantify the trade-off between rollout accuracy and computational cost. We further assess the model generalizability by testing a model trained on $\sim$5000 grains on flows containing $\sim$10,000 particles. By combining grain-level message passing with information exchange on coarser graphs, HCG-GNN reproduces the granular flow physics at both particle- and bulk-scale, and it sheds light on more efficient particle-resolved surrogate modeling for rapid forward prediction and inverse analyses.

\section{Numerical modeling and dataset}

\begin{figure}[h!]
    \centering
    \includegraphics[width=1.0\linewidth]{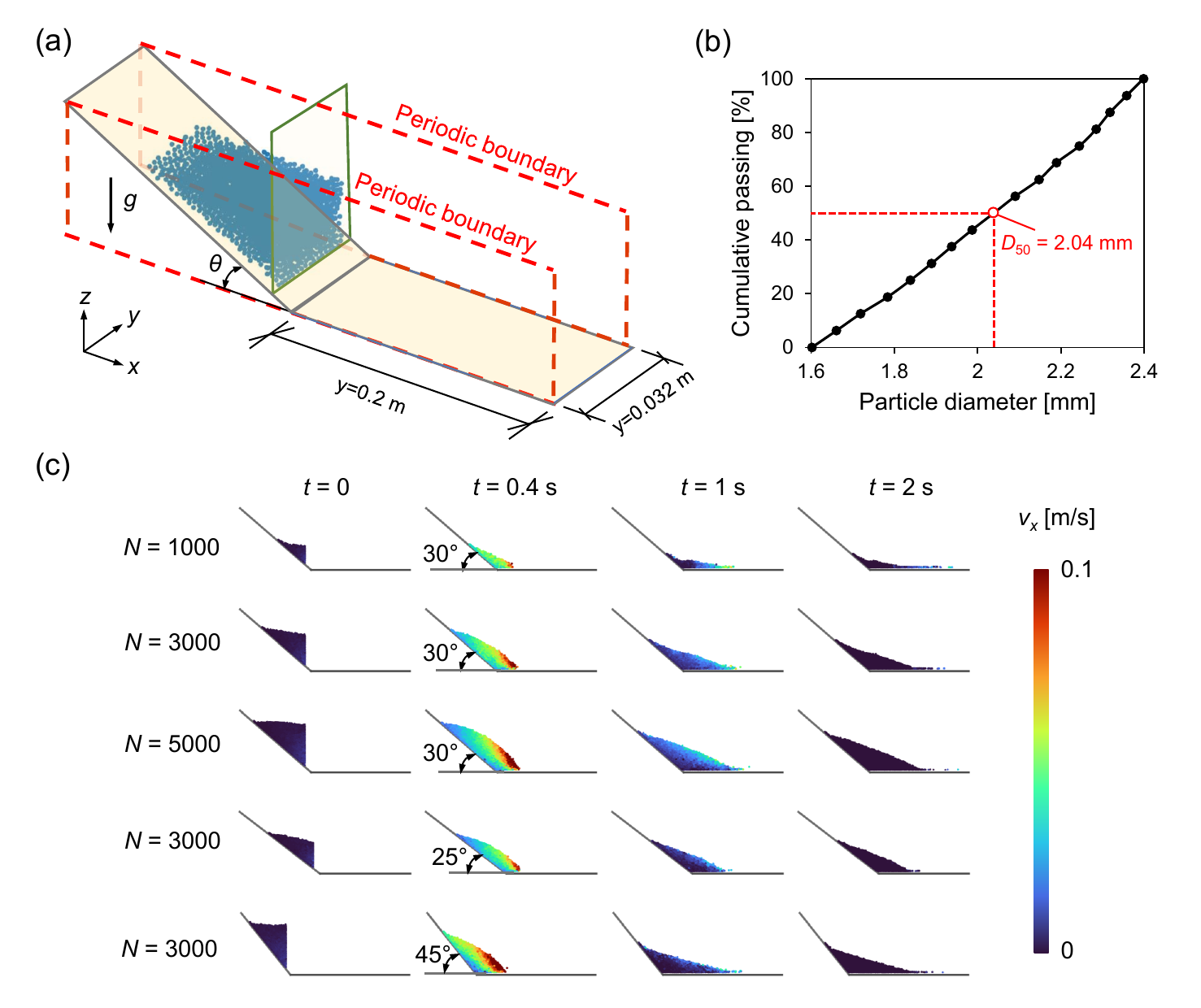}
    \caption{Discrete element model and simulation dataset for granular flow on inclined planes. (a) Schematic of the discrete element model, where particles are initialized, deposited on an inclined plane, and released by removing the vertical gate. Model geometry, coordinate system, inclination angle $\theta$, gravity direction, and boundary conditions are marked. (b) Particle size distribution used in the DEM simulations, where $D_{50}$ is the median particle diameter. (c) Representative DEM simulations with different particle numbers $N$ and inclination angles $\theta$. The evolution of the granular flow at $t$=0, 0.4, 1, and 2~s is presented with color-coded streamwise flow velocity.}
    \label{fig:DEM model and dataset}
\end{figure}

Figure~\ref{fig:DEM model and dataset}(a) presents the DEM configuration used to generate particle-resolved data for granular flow on an inclined plate connected to a horizontal plate. A vertical gate confined the particles on the inclined plate while they settled under gravity. Once a stable packing formed, the gate was removed, allowing the particles to flow down the inclined plate and onto the adjoining horizontal plate. The \(x\)-, \(y\)-, and \(z\)-axes denote the streamwise, transverse, and vertical directions, respectively. The domain had a transverse width of 0.032~m, with periodic boundary conditions applied to the two lateral boundaries in the \(y\)-direction~\cite{chen2026collapse}.

The granular material consisted of polydisperse spheres with a median diameter $D_{50}$ of approximately 2~mm. Figure~\ref{fig:DEM model and dataset}(b) shows the particle size distribution, and Table~\ref{tab:dem_parameters} lists the particle density, normal and tangential contact stiffnesses, and friction coefficients for particle--particle and particle--wall contacts. Each DEM simulation lasted 2~s, with particle states recorded at intervals of $\Delta t=0.02$~s, resulting in 100 frames for each simulation case. Figure~\ref{fig:DEM model and dataset}(c) illustrates the streamwise velocity $v_x$ throughout the flow process for representative combinations of inclination angle \(\theta\) and particle count \(N\).

\begin{table}[h!]
\centering
\caption{DEM simulation parameters}
\label{tab:dem_parameters}
\begin{tabular}{lll}
\toprule
Category & Parameter & Value \\
\midrule
Particle properties & Density & $2.56~\mathrm{g/cm^3}$ \\
                    & $D_{50}$ & $\approx 2~\mathrm{mm}$ \\
\midrule
Contact properties  & Normal stiffness, $K_n$ & $1 \times 10^{5}~\mathrm{N/m}$ \\
                    & Tangential stiffness, $K_t$ & $4 \times 10^{4}~\mathrm{N/m}$ \\
                    & $\mu_p$ & $0.6$ \\
                    & $\mu_w$ & $0.4$ \\
\bottomrule
\end{tabular}
\end{table}

The dataset was partitioned by complete trajectory (Table~\ref{tab:dataset_split}). The training set consisted of 9 trajectories covering \(\theta=25^\circ\), \(30^\circ\), and \(35^\circ\) with \(N=1000\), \(3000\), and \(5000\) particles. The validation set contained 3 trajectories at the unseen slope angle \(\theta=40^\circ\), while the test set consisted of 6 trajectories at \(\theta=30^\circ\) and \(45^\circ\) with the same particle numbers. Different random seeds were used across the training, validation, and test sets to generate distinct initial particle packing and avoid overlap between the datasets. The \(30^\circ\) test cases assessed prediction for new packings at an angle represented in training, while the unseen \(45^\circ\) cases required extrapolation beyond the training range of inclinations. Separate cases containing \(7500\) and \(10{,}000\) particles were used to assess extrapolation in particle count.

\begin{table}[h!]
\centering
\caption{Dataset split across inclination angles, particle counts, and random seed.}
\label{tab:dataset_split}
\begin{tabular}{l c c p{7cm}}
\toprule
Split & Angles & Particle counts & Purpose \\
\midrule
Training
& $25^\circ$, $30^\circ$, $35^\circ$
& 1000, 3000, 5000
& Learning across angle and size \\

Validation
& $40^\circ$
& 1000, 3000, 5000
& Model selection and hyperparameter tuning \\

Test A
& $45^\circ$
& 1000, 3000, 5000
& Angle extrapolation \\

Test B
& $30^\circ$
& 1000, 3000, 5000
& Unseen initial-condition test \\
\bottomrule
\end{tabular}
\end{table}

\section{Methodology}
\subsection{Graph neural network}
\label{sec:Graph neural network}

\begin{figure}[h!]
    \centering
    \includegraphics[width=1\linewidth]{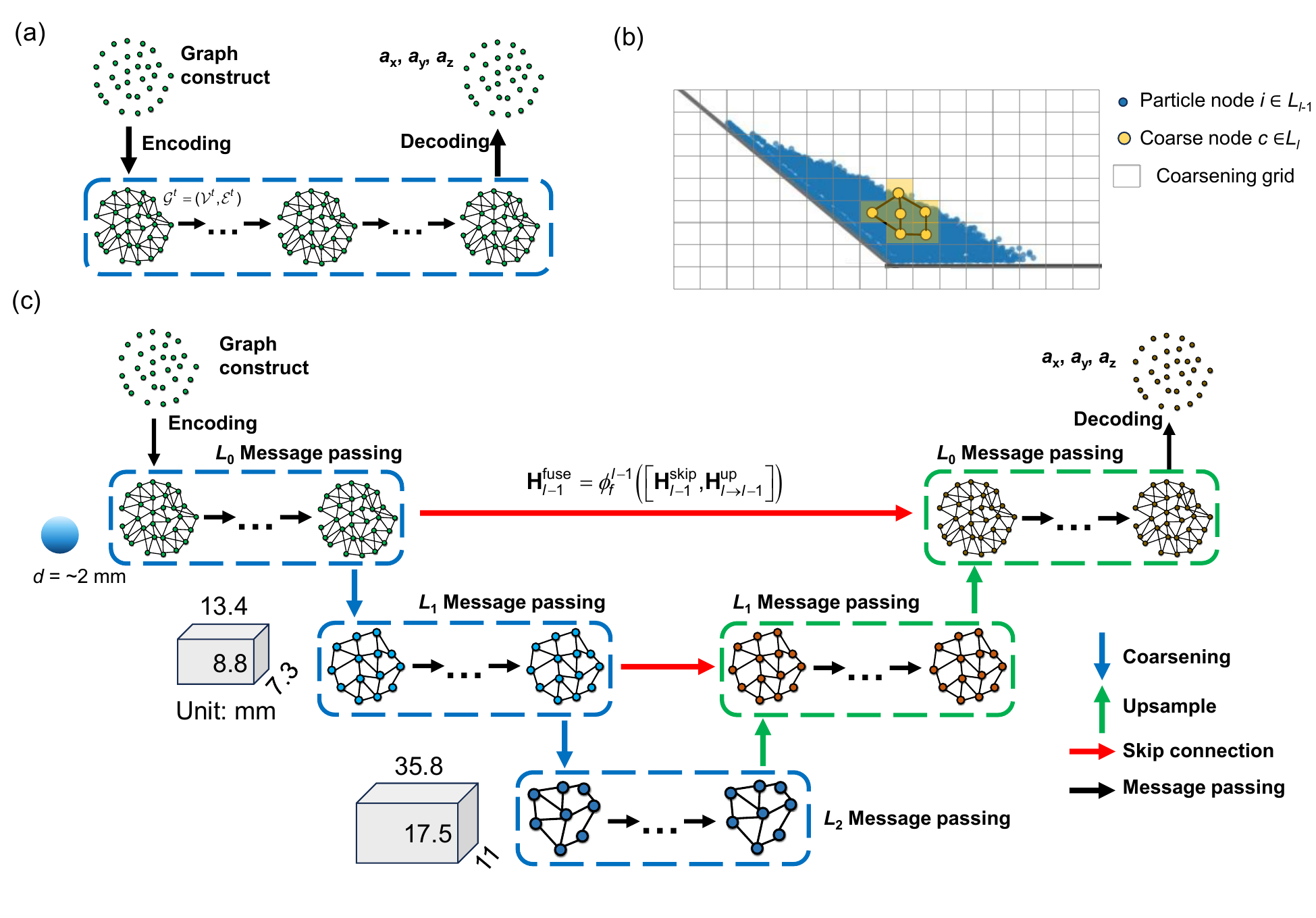}
    \caption{The architecture of Hierarchical Coarse-Grained Graph Neural Network (HCG-GNN). (a) Baseline GNN. The particle system is first converted into a graph, where particles are represented as nodes and local interactions are represented by edges. The graph is encoded into latent features, updated through repeated message passing, and decoded to the predicted particle acceleration components ($a_x$, $a_y$, $a_z$). (b) Coarse-grid representation of the granular system. Particle features are projected onto a structured spatial grid covering the granular domain. Each coarse cell aggregates information from the particles located inside the cell. (c) The proposed HCG-GNN architecture, which combines particle-level graph message passing with hierarchical coarse-grid feature propagation. Fine-scale particle features are first processed at level L0 and then coarsened to lower-resolution levels L1 and L2 to enlarge the receptive field. During decoding, coarse features are upsampled back to finer levels, while skip connections concatenate the original encoder features with the upsampled features at the corresponding decoder levels.}
    \label{fig:HCG-GNN framework}
\end{figure}

As illustrated in Figure~\ref{fig:HCG-GNN framework}(a), a graph neural network (GNN) represents a particle system at frame $t$ as a graph \(\mathcal{G}^{t}=(\mathcal{V}^{t},\mathcal{E}^{t})\) at time \(t\), where each node \(i\in\mathcal{V}^{t}\) represents a particle and each edge \((j,i)\in\mathcal{E}^{t}\) represents an interaction between two neighboring particles \cite{reiser2022gnnmaterials,schutt2018schnet}. The connectivity radius \(R_c\) defines the neighborhood of each particle: a particle \(j\) is a candidate neighbor of \(i\) when their center-to-center distance is at most \(R_c\). Particle properties and pairwise geometric quantities are encoded as node and edge features, respectively~\cite{xie2018crystal,fung2021benchmarking}. Here, node features describe velocity, particle size, and proximity to the boundaries, while edge features contain relative information between the connected particles, such as relative displacement, separation distance, and geometric overlap.

Particles gather information from neighboring and more distant particles through message passing over the graph \cite{kang2024messagepassing}. At message passing step \(m\), an edge message is first computed as
\(\mathbf{g}_{ij}^{m+1}=\phi_{e}(\mathbf{h}_{i}^{m},\mathbf{h}_{j}^{m},\mathbf{g}_{ij}^{m})\),
where \(\mathbf{h}_{i}^{m}\) and \(\mathbf{h}_{j}^{m}\) are the features of the receiver and sender particles, respectively, \(\mathbf{g}_{ij}^{m}\) represents the edge feature between them, and \(\phi_{e}\) is the edge update function. The incoming messages are then aggregated at each particle as \(\mathbf{m}_{i}^{m+1}=\sum_{j:(j,i)\in\mathcal{E}^{t}}\mathbf{g}_{ij}^{m+1}\), and the corresponding node feature is updated according to \(\mathbf{h}_{i}^{m+1}=\phi_{v}(\mathbf{h}_{i}^{m},\mathbf{m}_{i}^{m+1})\), where \(\phi_{v}\) is the node update function. Repeating this process allows information to propagate through the particle graph, and the resulting node representations can be used to predict particle level quantities such as acceleration \cite{unke2019physnet}. Further details of the conventional GNN formulation and implementation are provided in~\ref{app:conventional GNN}.

\subsection{Hierarchical Coarse-Grained Graph Neural Network (HCG-GNN)}
GNN's full-graph setup makes it expensive for large particle systems, while excessive coarsening may lose important spatial details. To balance efficiency and accuracy, we design a two-level grid-based coarsening hierarchy, where only limited message passing is performed on the fine graph, while most relational computation is shifted to the medium and coarse graph levels. 
The finest level, \(L_0\), retains one node per particle and represents the geometry of neighboring particle pairs. The two coarser levels, \(L_1\) and \(L_2\), exchange information among occupied spatial cells over progressively larger distances (Figure~\ref{fig:HCG-GNN framework}(c)). Direct message passing on \(L_0\) takes place both before coarsening and after the coarse features return to the particle graph. Thus, the hierarchy changes the spatial scale of information exchange while retaining individual particles as the units of motion prediction.

\subsubsection{Grid based coarsening}
\label{sec:grid based coarsening}

As shown in Figure~\ref{fig:HCG-GNN framework}(b), the computational domain is partitioned into regular grid cells, and these cells provide the spatial basis for graph coarsening. Each particle is assigned to a grid cell according to its position. The computational domain is defined as
\begin{align}
\Omega =
[x_{\min}, x_{\max}]
\times
[y_{\min}, y_{\max}]
\times
[z_{\min}, z_{\max}] .
\end{align}
where \(\Omega\) denotes the DEM computational domain, and \(x_{\min}\), \(x_{\max}\), \(y_{\min}\), \(y_{\max}\), \(z_{\min}\), and \(z_{\max}\) define the coordinate bounds of the domain.

The domain is discretized into structured grids with different resolutions:
\begin{align}
\mathbf{N}_{l}
=
\left(
N_{x,l},
N_{y,l},
N_{z,l}
\right).
\end{align}
where \(\mathbf{N}_{l}\) denotes the grid resolution at coarsening level \(l\), and \(N_{x,l}\), \(N_{y,l}\), and \(N_{z,l}\) are the numbers of grid cells along the \(x\)-, \(y\)-, and \(z\)-directions, respectively. 

At each coarsening level, a node from the preceding level is assigned to a grid cell according to its spatial position. Let \(\alpha_l(i)\) denote the grid cell assigned to node \(i\) during the transition from level \(L_{l-1}\) to \(L_l\), so the set of nodes assigned to an occupied cell \(c\) can be defined as:
\begin{align}
\mathcal{P}_{c}^{l}
=
\left\{
i \; \middle| \; \alpha_l(i)=c
\right\}.
\end{align}
Only cells containing at least one node are retained as nodes of the coarse graph. Empty cells are discarded and do not participate in subsequent graph computation. This operation allows the coarse graph to follow the evolving occupied region of the granular material while avoiding unnecessary computation in empty portions of the domain.

The features of all nodes assigned to the same occupied cell are aggregated by mean pooling. The pooled feature of cell \(c\) at level \(l\) is calculated as
\begin{align}
\bar{\mathbf{h}}_{c}^{l}
=
\frac{1}
{|\mathcal{P}_{c}^{l}|}
\sum_{i\in\mathcal{P}_{c}^{l}}
\mathbf{h}_{i}^{l-1},
\end{align}
where \(\mathbf{h}_{i}^{l-1}\) is the latent feature of node \(i\) at the preceding level $(l-1)$ and \(|\mathcal{P}_{c}^{l}|\) is the number of nodes assigned to cell \(c\). Mean pooling preserves the average latent representation but does not directly retain the number of nodes contained in the cell. To preserve this local concentration information, the occupancy of each cell is introduced as an additional feature, so the initial coarse node representation can be defined as
\begin{align}
\mathbf{h}_{c}^{l,0}
=
\phi_{p}^{l}
\left(
\left[
\bar{\mathbf{h}}_{c}^{l},
\log\left(1+|\mathcal{P}_{c}^{l}|\right)
\right]
\right),
\qquad l\in\{1,2\},
\end{align}
where \(\phi_{p}^{l}\) is a multilayer perceptron (MLP) used to transform the pooled features, and \([\cdot,\cdot]\) denotes feature concatenation. At \(L_1\), \(|\mathcal{P}_c^1|\) counts the particles assigned to cell \(c\). At \(L_2\), \(|\mathcal{P}_c^2|\) counts the occupied \(L_1\) cells assigned to the coarser cell. The logarithmic occupancy term distinguishes cells with similar mean features but different numbers of contributing nodes.

For each occupied grid cell \(c\) at coarsening level \(l\), its center coordinate is denoted by \(\mathbf{s}_{c}^{l}\) and used as the spatial position of the corresponding coarse node. Edges of the coarse graph are then constructed between occupied coarse nodes using a level-dependent connectivity radius. For each coarse level, the connectivity radius was defined based on the diagonal length of the corresponding grid cell as
\[
R_l=\eta\|\Delta\mathbf{d}_l\|_2,
\]
where \(\|\Delta\mathbf{d}_l\|_2\) is the cell diagonal and \(\eta\) is a connectivity factor. In this study, \(\eta=1.75\) was used to provide sufficient connectivity between neighboring occupied cells while maintaining a sparse coarse graph. Two occupied cells \(q\) and \(c\) are connected when
\begin{align}
(q,c)\in\mathcal{E}_{l}
\qquad
\mathrm{if}
\qquad
\left\|
\mathbf{s}_{c}^{l}
-
\mathbf{s}_{q}^{l}
\right\|_2
\leq R_l.
\end{align}
The corresponding coarse edge feature is defined as
\begin{align}
\mathbf{a}_{qc}^{l}
=
\left[
\frac{\mathbf{s}_{c}^{l}-\mathbf{s}_{q}^{l}}{R_l},
\left\|
\frac{\mathbf{s}_{c}^{l}-\mathbf{s}_{q}^{l}}{R_l}
\right\|_2
\right],
\end{align}
The edge feature contains the normalized relative displacement vector from the sender node \(q\) to the receiver node \(c\), together with the corresponding normalized Euclidean distance.

The same procedure is applied successively from \(L_0\) to \(L_1\) and from \(L_1\) to \(L_2\). In the first coarsening operation, particle features and particle positions are used to construct the \(L_1\) graph. In the second operation, the processed \(L_1\) node features and the centers of the occupied \(L_1\) cells are further pooled to construct the \(L_2\) graph. Importantly, the node-to-cell assignment relationships established during coarsening are retained and subsequently used to transfer coarse scale features back to the corresponding finer-level nodes during upsampling.

\subsubsection{Hierarchical message passing and feature transfer}

The proposed HCG-GNN propagates particle information through a hierarchical graph structure consisting of the particle level \(L_0\) and two coarsened levels \(L_1\) and \(L_2\), as illustrated in Figure~\ref{fig:HCG-GNN framework}(c). The overall architecture is inspired by U-Net, a widely used encoder--decoder architecture in the field of image recognition and image segmentation \cite{zhao2025pavement,ma2025crtransunet}. U-Net progressively extracts features at coarser spatial resolutions and subsequently transfers them back to finer resolutions while incorporating skip connections to preserve fine scale information. In the proposed framework, this multiscale concept is adapted to particle based graph representations through graph coarsening and hierarchical feature transfer. The conventional particle graph construction and message passing operations follow the formulation introduced in Section~\ref{sec:Graph neural network}. This section focuses on the input--output representation and the hierarchical information transfer operations that distinguish the proposed model from a conventional GNN.

\noindent \textbf{\textit{Input and output.}} 
For each training sample, an input sequence with a length of \(L=6\) consecutive particle-state frames is used. The position sequence of particle \(i\) is written as \(\{\mathbf{p}_{i}^{t-5},\mathbf{p}_{i}^{t-4},\ldots,\mathbf{p}_{i}^{t}\}\), where \(\mathbf{p}_{i}^{k}=[x_i^k,y_i^k,z_i^k]\) denotes the three-dimensional position of particle \(i\) at frame \(k\), and \(t\) denotes the current frame. Five particle velocities are obtained from the position differences between consecutive frames. The input feature of particle \(i\) is then defined as
\begin{align}
\mathbf{x}_{i}^{t}
=
\Big[
&
\widetilde{\mathbf{v}}_{i}^{t-L+2},
\ldots,
\widetilde{\mathbf{v}}_{i}^{t},
r_i,
\sin\theta,
\cos\theta,
\nonumber\\
&
\widetilde{d}_{i,c}^{\,t-L+1},
\ldots,
\widetilde{d}_{i,c}^{\,t},
\widetilde{d}_{i,b}^{\,t-L+1},
\ldots,
\widetilde{d}_{i,b}^{\,t}
\Big],
\end{align}
where \(\mathbf{x}_{i}^{t}\) denotes the input feature vector of particle \(i\) at time \(t\). 
The terms \(\widetilde{\mathbf{v}}_{i}^{t-L+2},\ldots,\widetilde{\mathbf{v}}_{i}^{t}\) represent the normalized velocity history of particle \(i\) over the input sequence, which describes its recent motion. 
The particle radius is denoted by \(r_i\), while \(\sin\theta\) and \(\cos\theta\) are used to represent the inclination angle \(\theta\). 
The terms \(\widetilde{d}_{i,c}^{\,t-L+1},\ldots,\widetilde{d}_{i,c}^{\,t}\) and \(\widetilde{d}_{i,b}^{\,t-L+1},\ldots,\widetilde{d}_{i,b}^{\,t}\) represent the normalized histories of the shortest distances from the particle center to the inclined plate and the horizontal plate, respectively. 
These distance histories describe the changing spatial relationship between the particle and the boundaries during the input sequence. 
For \(L=6\), the velocity history contributes 15 components, the particle radius contributes one component, the inclination angle contributes two components, and the two boundary distance histories contribute 12 components, resulting in a 30-dimensional node feature. The corresponding fine-scale edge feature is
\begin{align}
\mathbf{e}_{ij}^{t}
=
\left[
\frac{\Delta\mathbf{p}_{ij}^{t}}{R_c},
\frac{\left\|\Delta\mathbf{p}_{ij}^{t}\right\|_2}{R_c},
o_{ij}^{t}
\right],
\end{align}
where \(\Delta\mathbf{p}_{ij}^{t}=\mathbf{p}_{i}^{t}-\mathbf{p}_{j}^{t}\), \(R_c\) is the particle level connectivity radius, and \(o_{ij}^{t}\) is the normalized particle overlap. Thus, the edge feature contains three normalized relative displacement components, one normalized distance, and one overlap term, resulting in five dimensions.

Before message passing, the particle node and edge features are encoded into latent space using separate MLP,
\begin{align}
\mathbf{H}_{0}^{0}
&=
\phi_{v}^{\mathrm{enc}}(\mathbf{X}_{0}),
\\
\mathbf{G}_{0}^{0}
&=
\phi_{e}^{\mathrm{enc}}(\mathbf{E}_{0}),
\end{align}
where \(\mathbf{X}_{0}\) and \(\mathbf{E}_{0}\) are the node and edge feature matrices of the particle graph, respectively. The encoded features are then used for subsequent graph operations.

\noindent \textbf{\textit{Hierarchical feature transfer.}}
Message passing is first performed on the particle graph \(L_0\),
\begin{align}
\mathbf{H}_{0}^{\mathrm{mp}}
=
\operatorname{MP}_{0}^{K_0}
\left(
\mathbf{H}_{0}^{0},
\mathbf{G}_{0}^{0},
\mathcal{E}_{0}
\right),
\end{align}
where \(\mathbf{H}_{0}^{0}\) and \(\mathbf{G}_{0}^{0}\) denote the initial node and edge features of the particle graph, respectively, \(\mathcal{E}_{0}\) denotes its graph connectivity, and \(K_0\) is the number of particle level message passing steps. The resulting features \(\mathbf{H}_{0}^{\mathrm{mp}}\) contain the processed particle information and are retained for a later skip connection.

The processed features are then transferred successively to the two coarse levels using the same pooling and message passing operations. For \(l=1,2\), the hierarchical transfer is expressed as
\begin{align}
\mathbf{H}_{l}^{0}
&=
\operatorname{Pool}_{l-1\rightarrow l}
\left(
\mathbf{H}_{l-1}^{\mathrm{mp}}
\right),
\\
\mathbf{H}_{l}^{\mathrm{mp}}
&=
\operatorname{MP}_{l}^{K_l}
\left(
\mathbf{H}_{l}^{0},
\mathbf{G}_{l}^{0},
\mathcal{E}_{l}
\right).
\end{align}
where \(\operatorname{Pool}_{l-1\rightarrow l}\) denotes the grid based coarsening operation from level \(L_{l-1}\) to level \(L_l\), as described in Section~\ref{sec:grid based coarsening}, and \(\mathbf{H}_{l}^{0}\) denotes the resulting initial node features at the coarse level. Only occupied grid cells are retained as coarse nodes. The quantities \(\mathbf{G}_{l}^{0}\) and \(\mathcal{E}_{l}\) denote the corresponding edge features and graph connectivity, respectively, while \(K_l\) is the number of message passing steps at level \(L_l\). The resulting \(\mathbf{H}_{l}^{\mathrm{mp}}\) represents the processed node features after coarse level message passing.

For \(l=1\), pooling maps the processed particle features from \(L_0\) to the occupied cells of \(L_1\). For \(l=2\), the same operation is applied to the processed \(L_1\) nodes and their cell center positions, rather than directly to the original particles. Thus, information is progressively transferred through \(L_0\rightarrow L_1\rightarrow L_2\). Since each coarse node represents a larger spatial region while the numbers of active nodes and edges decrease after pooling, coarse level message passing enables information to propagate over larger physical distances with reduced graph complexity.

\noindent \textbf{\textit{Upsampling and skip connections.}}
After message passing at the coarsest level \(L_2\), the coarse scale information is progressively transferred back through \(L_2\rightarrow L_1\rightarrow L_0\). For two adjacent graph levels \(L_l\) and \(L_{l-1}\), the upsampling operation is generally written as
\begin{align}
\mathbf{H}_{l\rightarrow l-1}^{\mathrm{up}}
=
\operatorname{Up}_{l\rightarrow l-1}
\left(
\mathbf{H}_{l}
\right),
\qquad
l\in\{2,1\},
\end{align}
where \(\mathbf{H}_{l}\) denotes the feature representation at the coarser level \(L_l\), \(\operatorname{Up}_{l\rightarrow l-1}(\cdot)\) denotes the upsampling operation from \(L_l\) to \(L_{l-1}\), and \(\mathbf{H}_{l\rightarrow l-1}^{\mathrm{up}}\) represents the resulting features transferred to the finer level. The node-to-cell assignment relationships recorded during the corresponding pooling operation are used for this feature transfer, so that each finer-level node receives the feature of its associated coarse node.

At each finer level, the upsampled features are combined with the features retained before the corresponding pooling operation through a skip connection,
\begin{align}
\mathbf{H}_{l-1}^{\mathrm{fuse}}
=
\phi_{f}^{l-1}
\left(
\left[
\mathbf{H}_{l-1}^{\mathrm{skip}},
\mathbf{H}_{l\rightarrow l-1}^{\mathrm{up}}
\right]
\right),
\qquad
l\in\{2,1\},
\end{align}
where \(\mathbf{H}_{l-1}^{\mathrm{skip}}\) denotes the finer-level feature retained before coarsening, \([\cdot,\cdot]\) denotes feature concatenation, and \(\phi_{f}^{l-1}\) is the corresponding feature fusion MLP. Specifically, the \(L_2\) features are first transferred to \(L_1\) and combined with the retained \(L_1\) features. The resulting \(L_1\) representation is then transferred to \(L_0\) and combined with the retained particle level features. In this way, the long range information obtained from the coarse graphs is progressively integrated with the finer-scale information preserved through the skip connections.

After the hierarchical features are transferred back to the particle level, message passing is performed again on the fine graph,
\begin{align}
\mathbf{H}_{0}^{\mathrm{out}}
=
\operatorname{MP}_{0}^{K_0^{u}}
\left(
\mathbf{H}_{0}^{\mathrm{fuse}},
\mathbf{G}_{0}^{0},
\mathcal{E}_{0}
\right),
\end{align}
where \(K_0^{u}\) denotes the number of message passing steps at the particle level after feature fusion, and \(\mathbf{H}_{0}^{\mathrm{out}}\) is the final particle level latent representation. This operation allows the transferred coarse scale information to interact again with the local particle neighborhood before acceleration prediction.

\noindent \textbf{\textit{Output decoder.}}
Finally, the particle level latent features are decoded into the normalized particle accelerations,
\begin{align}
\widehat{\mathbf{a}}_{i,n}^{t}
=
\phi_{\mathrm{dec}}
\left(
\mathbf{h}_{i}^{0,\mathrm{out}}
\right),
\end{align}
where \(\mathbf{h}_{i}^{0,\mathrm{out}}\) denotes the final latent feature of particle \(i\) at level \(L_0\), \(\phi_{\mathrm{dec}}\) is the output decoder, and \(\widehat{\mathbf{a}}_{i,n}^{t}\in\mathbb{R}^{3}\) represents the predicted normalized acceleration of particle \(i\) in the \(x\)-, \(y\)-, and \(z\)-directions. The predicted acceleration is subsequently transformed back to the physical scale and used to update the particle velocity and position through the time integration scheme described in the following section.

\subsection{Physical time integration}
\label{sec:Physical time integration}

The network predicts the normalized acceleration of each particle rather than its next position directly. The predicted acceleration is first transformed back to physical units and then used to update the particle velocity and position through time integration. For particle \(i\), the update procedure is written as
\begin{align}
\widehat{\mathbf{a}}_{i}^{t}
&=
\widehat{\mathbf{a}}_{i,n}^{t}
\odot
\boldsymbol{\sigma}_{a}
+
\boldsymbol{\mu}_{a},
\\
\widehat{\mathbf{v}}_{i}^{t+1}
&=
\mathbf{v}_{i}^{t}
+
\widehat{\mathbf{a}}_{i}^{t}\Delta t,
\\
\widehat{\mathbf{p}}_{i}^{t+1}
&=
\mathbf{p}_{i}^{t}
+
\widehat{\mathbf{v}}_{i}^{t+1}\Delta t,
\end{align}
where \(\widehat{\mathbf{a}}_{i,n}^{t}\) and \(\widehat{\mathbf{a}}_{i}^{t}\) are the predicted accelerations in normalized and physical units, respectively. The vectors \(\boldsymbol{\mu}_{a}\) and \(\boldsymbol{\sigma}_{a}\) contain the componentwise means and standard deviations of acceleration in the training dataset, and \(\odot\) denotes elementwise multiplication. The update uses a semi-implicit Euler step of duration \(\Delta t\): the predicted acceleration first advances the velocity, and this updated velocity then advances the position. Each application of the network and integration scheme advances the particle configuration by one recorded frame.
and \(\Delta t\) is the time interval between two consecutive frames. Thus, the neural network predicts only the particle acceleration, while the velocity and position are updated through deterministic time integration.

To improve robustness to accumulated prediction errors during rollout, random-walk noise is introduced into the input position sequence during training. For a clean sequence \(\mathbf{P}_{i}^{t-L+1:t}=[\mathbf{p}_{i}^{t-L+1},\ldots,\mathbf{p}_{i}^{t}]\), the noisy sequence is written as \(\widetilde{\mathbf{P}}_{i}^{t-L+1:t}=\mathbf{P}_{i}^{t-L+1:t}+\boldsymbol{\mathcal{E}}_{i}^{t-L+1:t}\). The perturbation is generated by sampling random increments \(\boldsymbol{\eta}_{i}^{k}\sim\mathcal{N}(\mathbf{0},\sigma_n^2\mathbf{I})\) and accumulating them over the input sequence to produce temporally correlated position noise. This treatment introduces gradually accumulated perturbations rather than independent noise at each frame, thereby better representing the error accumulation encountered during autoregressive prediction.

During autoregressive rollout, the first \(L=6\) particle state frames, \(\{\mathbf{P}^{t-5},\ldots,\mathbf{P}^{t}\}\), are provided as the initial input sequence. The model predicts the particle acceleration at frame \(t\), and the corresponding position \(\widehat{\mathbf{P}}^{t+1}\) is obtained through the above time integration scheme. The predicted configuration is then appended to the sequence while the earliest frame is removed, forming the next input window \(\{\mathbf{P}^{t-4},\ldots,\mathbf{P}^{t},\widehat{\mathbf{P}}^{t+1}\}\). This sliding-window procedure is repeated recursively to generate the subsequent particle configurations.

\subsection{Training strategy}

Training samples were extracted as sliding windows from the 9 DEM trajectories assigned to the training set in Table~\ref{tab:dataset_split}. Each sample paired 6 consecutive input frames with the target particle accelerations for the transition to the following frame. The model parameters were optimized using the Adam optimizer, and the main training and model hyperparameters are summarized in Table~\ref{tab:gnn_unet_hyperparameters}.

The training objective combines acceleration, velocity, and position errors and is defined as
\begin{align}
\mathcal{L}
=
\lambda_a\mathcal{L}_a
+
\lambda_v\mathcal{L}_v
+
\lambda_p\mathcal{L}_p,
\end{align}
where $\mathcal{L}_a$, $\mathcal{L}_v$, and $\mathcal{L}_p$ denote the mean squared errors (MSEs) of acceleration, velocity, and position, respectively, and $\lambda_a$, $\lambda_v$, and $\lambda_p$ are their corresponding weighting coefficients. The acceleration loss is calculated from the difference between the predicted and target normalized particle accelerations. The predicted acceleration is then transformed to the physical scale and integrated to obtain the next-frame particle velocity and position. The velocity and position losses are calculated from the differences between these one-step predictions and the corresponding DEM values. Appropriate normalization is applied to the velocity and position errors to maintain comparable numerical scales among the three loss terms. In this study, the loss weights were set to \(\lambda_a=1.0\), \(\lambda_v=0.25\), and \(\lambda_p=0.25\).

\begin{table}[h!]
\centering
\caption{HCG-GNN training hyperparameters.}
\label{tab:gnn_unet_hyperparameters}
\begin{tabular}{ll}
\toprule
Hyperparameter & Value \\
\midrule
Input sequence length & \(6\) \\
Optimizer & Adam \\
Initial learning rate & \(1\times10^{-4}\) \\
Learning-rate decay & \(0.99998\) \\
Minimum learning rate & \(1\times10^{-7}\) \\
Noise standard deviation & \(2\times10^{-4}\) \\
Loss weights, \((\lambda_a,\lambda_v,\lambda_p)\) & \((1.0,0.25,0.25)\) \\
Connectivity radius, \(R_c\) & \(0.9\) \\
Maximum neighbors & \(32\) \\
Latent dimension & \(256\) \\
Hidden dimension & \(256\) \\
\(L_0\) message passing steps before pooling & \(1\) \\
\(L_1\) message passing steps & \(2\) \\
\(L_2\) message passing steps & \(1\) \\
\(L_1\) message passing steps after upsampling & \(2\) \\
\(L_0\) message passing steps after upsampling & \(1\) \\
\(L_1\) grid shape & \((32,6,16)\) \\
\(L_2\) grid shape & \((12,4,8)\) \\
\bottomrule
\end{tabular}
\end{table}

The one-step training objective measures prediction error for a single update. During autoregressive rollout, an error in particle position also affects the geometry used in subsequent updates and can accumulate over the trajectory. Therefore, we evaluated model checkpoints on complete validation rollouts using a composite score. The score combines particle trajectory errors with measures of the final flow configuration, bulk flow characteristics, and boundary penetration.

The position error over the rollout trajectory is quantified by the mean squared error (MSE),
\begin{align}
E_{\mathrm{MSE}}
=
\frac{1}{3TN}
\sum_{t=1}^{T}
\sum_{i=1}^{N}
\left\|
\boldsymbol{\delta}
\left(
\widehat{\mathbf{p}}_{i}^{t},
\mathbf{p}_{i}^{t}
\right)
\right\|_{2}^{2},
\end{align}
where \(T\) is the rollout horizon, \(N\) is the number of particles, \(\widehat{\mathbf{p}}_{i}^{t}\) and \(\mathbf{p}_{i}^{t}\) denote the predicted and DEM particle positions, respectively, and $\boldsymbol{\delta}(\cdot)$ denotes the difference between the predicted and DEM particle positions.

To evaluate the rollout more comprehensively, the composite rollout score is defined as
\begin{align}
S_{\mathrm{rollout}}
={}&
E_{\mathrm{MSE}}
+2.5E_{\mathrm{final}}
+1.5E_{\mathrm{P95}}
+E_{\mathrm{front}}
\nonumber\\
&+0.75E_{\mathrm{centroid}}
+0.75E_{\mathrm{number}}
+\left(1-D_{\mathrm{side}}\right)
+6P_{\mathrm{pen}},
\end{align}
where \(E_{\mathrm{final}}\) is the particle position MSE at the final rollout step, \(E_{\mathrm{P95}}\) is the 95th percentile of the particle position error at the final step, \(E_{\mathrm{front}}\) and \(E_{\mathrm{centroid}}\) measure the errors in the predicted flow front position and particle cloud centroid, respectively, \(E_{\mathrm{number}}\) represents the discrepancy in the streamwise particle number distribution, and \(D_{\mathrm{side}}\) is the Dice coefficient measuring the similarity between the predicted and DEM side view flow morphologies. The boundary penetration penalty is defined as
\begin{align}
P_{\mathrm{pen}}
=
P_{\mathrm{mean}}
+10P_{\mathrm{max}}
+4P_{\mathrm{frac}},
\end{align}
where \(P_{\mathrm{mean}}\), \(P_{\mathrm{max}}\), and \(P_{\mathrm{frac}}\) denote the mean penetration depth, maximum penetration depth, and fraction of particles penetrating the boundaries, respectively. A smaller \(S_{\mathrm{rollout}}\) therefore indicates better overall agreement between the predicted and DEM rollouts in terms of both particle level accuracy and global flow behavior. Rollout validation was performed every 100 training steps to monitor the long-term predictive performance of the model. The checkpoint with the lowest composite rollout score was selected as the best-performing model.

\section{Results}

\begin{figure}[h!]
    \centering
    \includegraphics[width=1\linewidth]{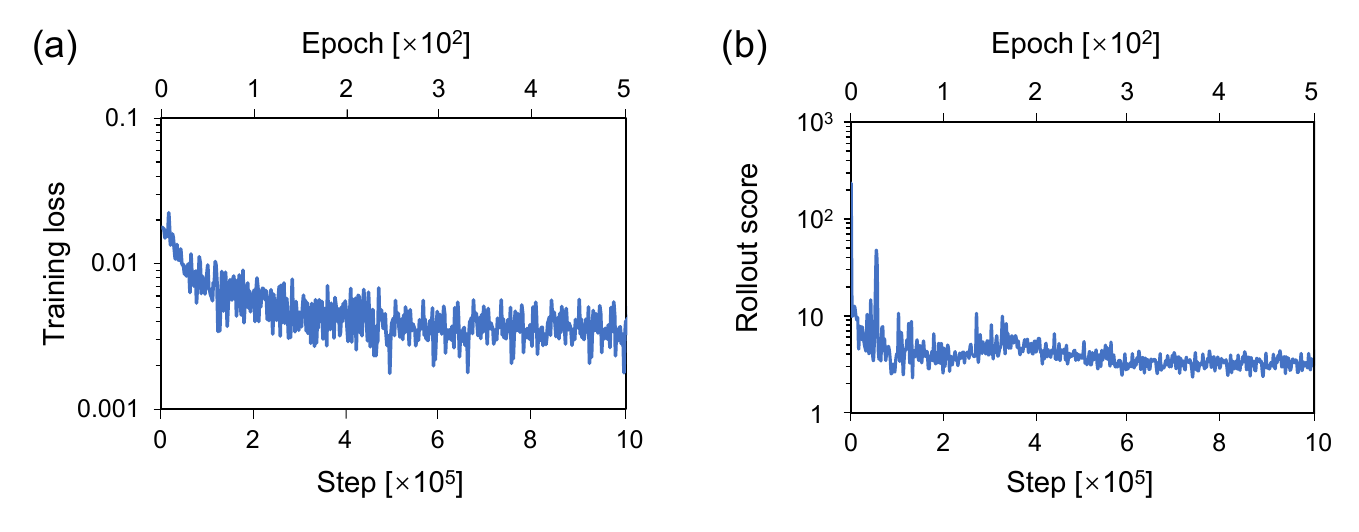}
    \caption{Training and rollout performance of the proposed HCG-GNN model. (a) Evolution of the training loss with training step/epoch. (b) Evolution of the rollout score during training.}
    \label{fig:Training loss and rollout loss}
\end{figure}

Figure~\ref{fig:Training loss and rollout loss} compares the evolution of the one-step training loss (Figure~\ref{fig:Training loss and rollout loss}(a)) with that of the validation score obtained from autoregressive rollouts (Figure~\ref{fig:Training loss and rollout loss}(b)), which reflects the accumulated error. The training loss decreases most rapidly during early optimization and subsequently fluctuates around 0.003--0.004. The rollout score also decreases sharply, although its early fluctuations are more pronounced since each predicted configuration becomes an input to subsequent updates. At later epochs, both measures vary over a narrower range, indicating that the improvement in one-step prediction is accompanied by lower accumulated error on the validation particle trajectories.

\subsection{Particle-level kinematics}

\begin{figure}[h!]
    \centering
    \includegraphics[width=\linewidth]{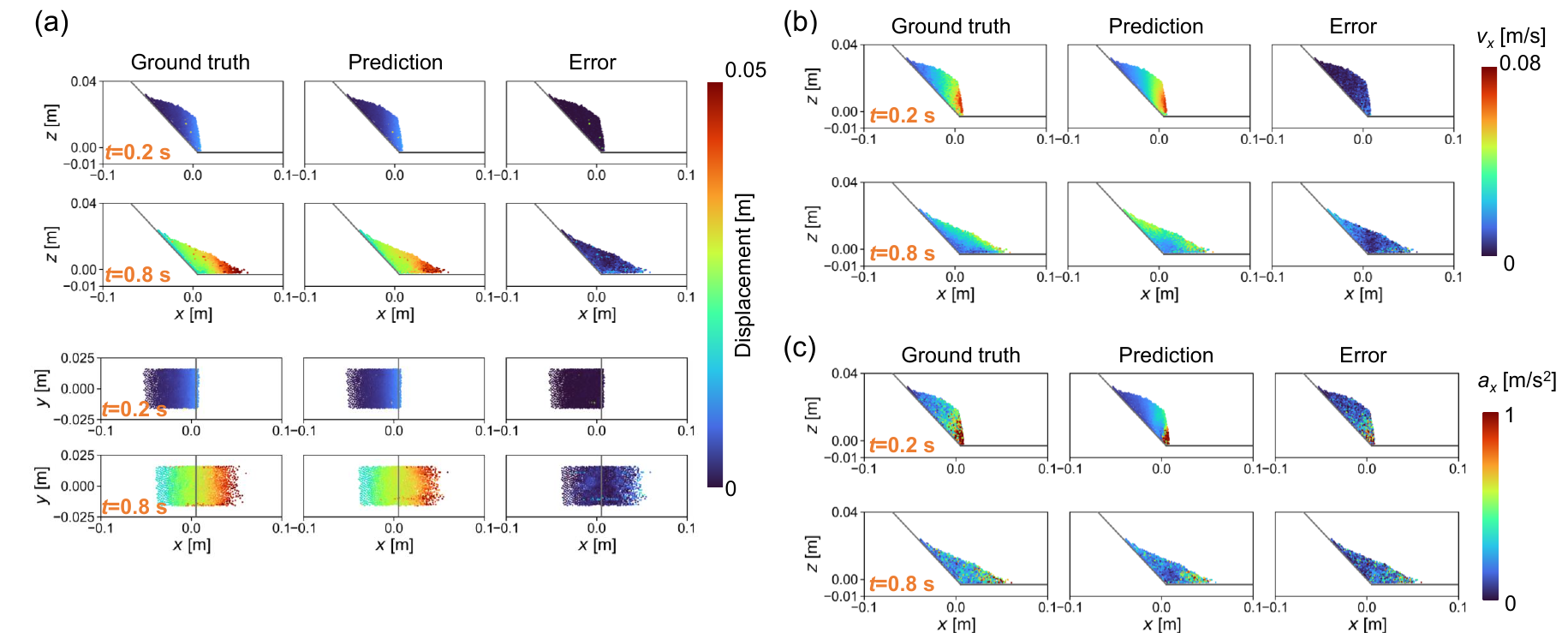}
    \caption{Particle-scale kinematic comparison of displacement, velocity, and acceleration between the DEM ground truth and HCG-GNN prediction. (a) Side- and top-view displacement magnitude distributions at \(t=0.2\)~s and \(0.8\)~s. (b,c) Distributions of streamwise velocity magnitude \(|v_x|\) and acceleration magnitude \(|a_x|\). For each quantity, the DEM ground truth, HCG-GNN prediction, and corresponding error are shown side by side.}
    \label{fig:displacement_velocity_acceleration}
\end{figure}

Figure~\ref{fig:displacement_velocity_acceleration} compares particle displacement, streamwise speed, and streamwise acceleration magnitude during the early release ($t=0.2$~s) and subsequent spreading ($t=0.8$~s). 
In both the side and top views, HCG-GNN reproduces the increase in displacement from the rear of the granular mass toward the advancing front. By $t=0.8$~s, the particles that have traveled farthest occupy the leading portion of the mass on the horizontal bed. The predicted streamwise speed also follows the DEM spatial pattern, including the region of faster motion near the front during early release. The acceleration field exhibits greater particle-to-particle variability in DEM than in the prediction, with the most visible local discrepancies near the front at $t=0.2$~s. This indicates that the model captures the accumulated downslope motion more closely than the short-lived acceleration variations of individual grains at the front. The local acceleration discrepancies occurres near the flow front, where frequent collisions, loss of contact, and reestablishment of contact lead to instantaneous particle acceleration fluctuations. However, the variations become smaller in velocity and displacement.
Overall, HCG-GNN reproduces the evolution of particle displacement, velocity, and acceleration during autoregressive rollout with high fidelity, despite localized errors in the instantaneous acceleration. Additional displacement comparisons for different inclination angles, particle counts, and particle initializations are provided in \ref{app:Additional granular flow prediction results}.

\subsection{Bulk flow evolution}
\begin{figure}[h!]
    \centering
    \includegraphics[width=1\linewidth]{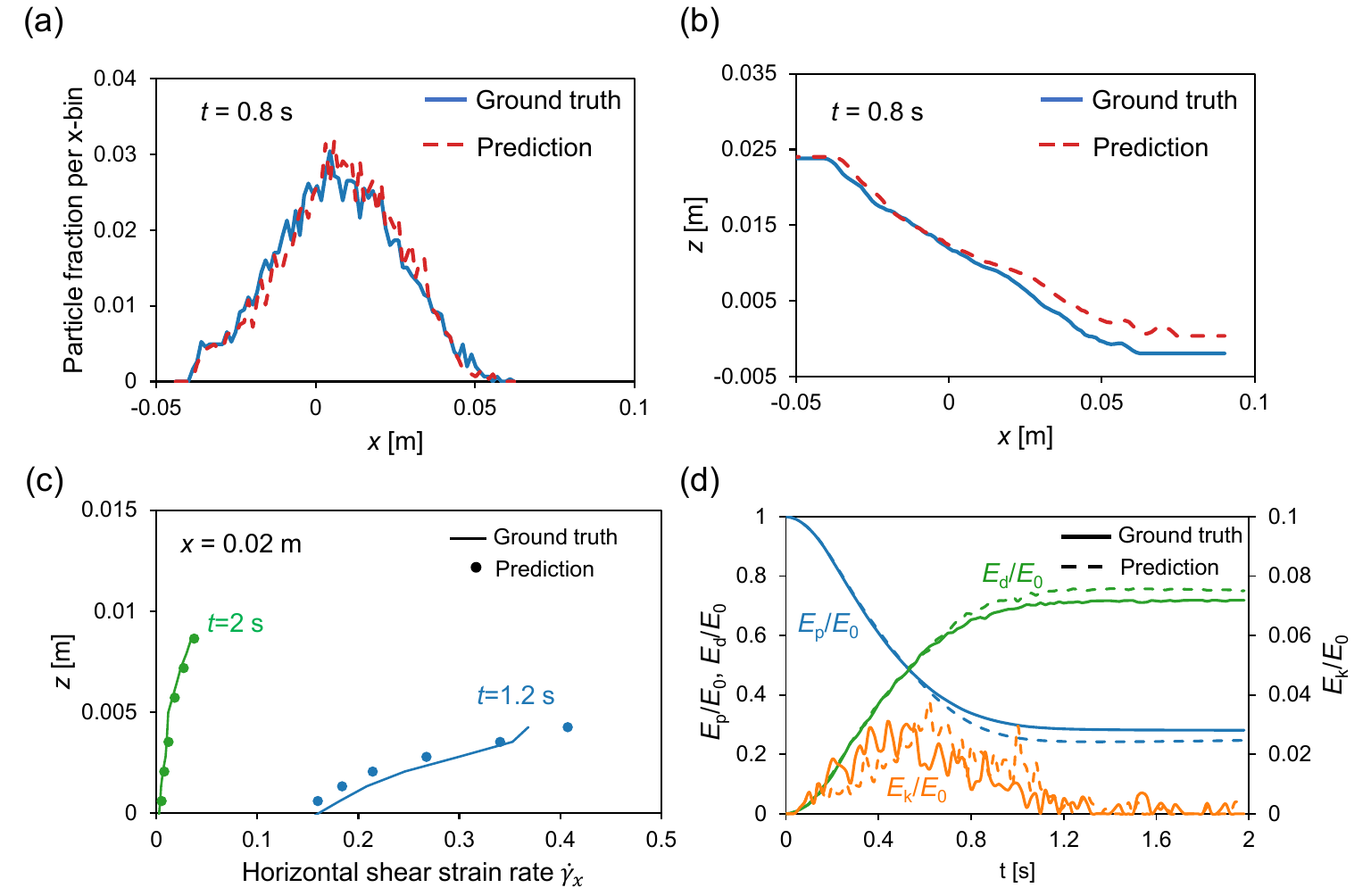}
    \caption{Comparison of bulk granular flow and internal kinematics between DEM ground truth and HCG-GNN rollouts. (a) Particle number fraction in each streamwise bin at $t$ = 0.8~s. (b) Free-surface elevation at the same time. (c) Horizontal shear strain rate profiles at $x$ = 0.02 m for $t=1.2$ and 2~s. (d) Gravitational potential energy $E_p$, translational kinetic energy $E_k$, and dissipated energy $E_d$, normalized by the initial mechanical energy $E_0$.}
    \label{fig:granular_flow_physics_evaluation}
\end{figure}

Figure~\ref{fig:granular_flow_physics_evaluation}(a) and (b) compare the streamwise distribution of particles and the resulting free-surface profile at $t=0.8$~s. The quantities are defined in \ref{app:physical_metrics}. Both the DEM and predicted particle number fractions (Figure~\ref{fig:granular_flow_physics_evaluation}(a)) reach a maximum within the first 0.01~m downstream of \(x=0\), with peak fractions of approximately 0.029 and 0.032, respectively. The predicted surface elevation (Figure~\ref{fig:granular_flow_physics_evaluation}(b)) closely follows the DEM profile over the rear and middle portions of the mass, decreasing from approximately 0.024~m at the rear. The profiles diverge toward the front, where HCG-GNN places the surface approximately 2~mm above the DEM surface. This difference corresponds to about 1 median grain diameter and is most apparent where the flowing layer is thin. Overall, the model can well reproduce the streamwise particle distribution while slightly overestimating the surface elevation near the flow front.

Figure~\ref{fig:granular_flow_physics_evaluation}(c) evaluates the internal motion at \(x=0.02\)~m using the horizontal shear strain rate profile. At \(t=1.2\)~s, the shear strain rate increases with elevation, and the HCG-GNN prediction remains close to the DEM curve, indicating that the model captures the depth-dependent velocity gradient inside the flowing layer. At \(t=2.0\)~s, both profiles lie close to zero throughout the sampled depth, consistent with the reduced shearing as the material approaches deposition. This comparison suggests that the learned simulator can reproduce not only the external flow profile but also the internal deformation.

Figure~\ref{fig:granular_flow_physics_evaluation}(d) compares the normalized energy evolution obtained from the DEM and HCG-GNN rollouts. The evolution of the potential energy \(E_p\), kinetic energy \(E_k\), and dissipated energy \(E_d\) follows the energy balance \(E_p+E_k+E_d=E_0\), where \(E_0\) is the initial mechanical energy, and all energy components are normalized by \(E_0\). The potential-energy histories closely follow one another as the mass moves downslope. The kinetic-energy histories show some differences during the first second of motion: the predicted peak is a little higher and occurs later than the DEM peak. These discrepancies are associated with the rapidly varying particle interactions and motion during this stage. As the flow subsequently decelerates, the kinetic energy decreases and the predicted energy components gradually approach the DEM results.

Taken together, the rollout comparisons demonstrate that the proposed HCG-GNN model can reproduce key physical characteristics of DEM granular flow, including downslope runout, lateral spreading, deposit profile, internal shear rate, and energy conversion. The main discrepancies occur near the moving front and during the most dynamic acceleration stage, where small one-step errors are amplified through autoregressive integration. Despite these local differences, the predicted trajectories remain physically stable and closely follow the DEM evolution, supporting the use of the model as a particle-resolved surrogate for quick granular flow prediction.

\section{Discussion}

\subsection{Spatial reach of message passing}

In a conventional particle graph with connectivity radius $r$, each message-passing step transfers information across at most one edge. After \(K\) steps, the distance between the endpoints of a propagation path is bounded by \(K r\), with the actual distance depending on particle connectivity along the path. HCG-GNN extends the distance covered by individual updates by grouping nearby particles into occupied cells and exchanging information between cell centers at successively coarser levels. In the present model, \(r=9.00\)~mm is the particle-level cutoff radius used to construct graph edges. 

At each coarse level \(L_l\) in HCG-GNN, the connection radius \(R_l\) is set to \(1.75\) times the cell diagonal \(D_{\mathrm{cell}}\), as defined in Section~\ref{sec:grid based coarsening}. The resulting radii at \(L_1\) and \(L_2\) are \(3.43 r\) and \(8.04 r\), respectively (Table~\ref{tab:coarse_graph_hops}). This indicates that a single update at \(L_2\) can connect occupied regions whose centers are separated by as much as eight particle-level cutoff radii. The distance over which information ultimately influences an individual grain also depends on the occupied-cell connectivity, the sequence of updates across levels, and the learned messages.

\begin{table}[h!]
\centering
\small
\caption{Quantitative comparison of information-propagation scales across graph levels}
\label{tab:coarse_graph_hops}

\begin{tabularx}{\textwidth}{
>{\raggedright\arraybackslash}p{0.15\textwidth}
>{\centering\arraybackslash}p{0.15\textwidth}
>{\centering\arraybackslash}p{0.27\textwidth}
>{\centering\arraybackslash}p{0.10\textwidth}
>{\centering\arraybackslash}p{0.08\textwidth}
>{\centering\arraybackslash}p{0.13\textwidth}
}
\toprule
Graph level
& Grid shape
& Cell size [mm]
& \(D_{\mathrm{cell}}\) [mm]
& \(R_l\) [mm]
& \(R_l/r\) \\
\midrule

Particle graph
& --
& --
& --
& 9.00
& 1.00 \\

\(L_1\)
& \(32\times6\times16\)
& \(13.44\times7.33\times8.75\)
& 17.63
& 30.86
& 3.43 \\

\(L_2\)
& \(12\times4\times8\)
& \(35.83\times11.00\times17.50\)
& 41.37
& 72.39
& 8.04 \\

\bottomrule
\end{tabularx}

\vspace{2pt}
\begin{minipage}{\textwidth}
\footnotesize
\textit{Note:}
\(D_{\mathrm{cell}}\) denotes the spatial diagonal length of a three-dimensional grid cell.
\(R_l\) denotes the connection radius used to construct the coarse graph at level \(L_l\), defined as \(R_l=1.75D_{\mathrm{cell}}\).
The particle-level connection radius is denoted by \(r\).
\end{minipage}
\end{table}

\subsection{Training cost relative to the particle-level GNN}

The two-level HCG-GNN and the conventional GNN with 10 particle-level message-passing steps were each trained for 500 epochs on the same NVIDIA H100 GPU with \(96~\mathrm{GB}\) of memory using a batch size of 2. As summarized in Table~\ref{tab:model_cost_comparison}, HCG-GNN reduced the trainable parameter count from 4.82 million to 3.24 million (32.7\%) and GPU memory usage from 40.271 to 21.659~GB (46.2\%), representing a saving of \(18.612~\mathrm{GB}\). This means that the hierarchical model requires only slightly more than half of the GPU memory consumed by the conventional fine-scale GNN under the tested training configuration. Additionally, HCG-GNN also decreased the mean time per epoch from 100.65 to 15.55~s, corresponding to a \(6.47\)-fold training speedup and a saving of approximately 11 h 49 min over 500 epochs.

The reduction in training time is substantially larger than the reduction in parameter count. The conventional GNN performs all 10 updates on the particle graph, whereas the two-level HCG-GNN performs 2 updates on that graph and 5 on graphs of occupied cells. This allocation reduces the number of times the full set of particle neighborhoods must be processed. Although pooling, upsampling, and skip feature fusion introduce additional operations, their computational cost is considerably lower than that of deep message passing on the full particle graph. The reduced memory requirement may also permit larger batch sizes or particle systems.

\begin{table}[h]
\centering
\caption{Comparison of model parameters and computational cost.}
\label{tab:model_cost_comparison}
\setlength{\tabcolsep}{4pt}
\begin{tabular}{
>{\centering\arraybackslash}p{2.3cm}
>{\centering\arraybackslash}p{2.3cm}
>{\centering\arraybackslash}p{2.2cm}
>{\centering\arraybackslash}p{2.0cm}
>{\centering\arraybackslash}p{2.4cm}
>{\centering\arraybackslash}p{1.8cm}
}
\toprule
Model
& \shortstack{Trainable\\parameters}
& \shortstack{GPU memory\\usage (GB)}
& \shortstack{Time per\\epoch (s)}
& \shortstack{Training time\\(500 epochs)}
& \shortstack{Training\\loss} \\
\midrule

10 MP GNN
& \(4{,}816{,}643\)
& \(40.27\)
& \(100.65\)
& \(13{:}58{:}45\)
& \(1.7\times10^{-3}\) \\

HCG-GNN
& \(3{,}240{,}195\)
& \(21.66\)
& \(15.55\)
& \(2{:}09{:}37\)
& \(1.5\times10^{-3}\) \\

\bottomrule
\end{tabular}
\end{table}


\subsection{Scalability to larger particle systems}

\begin{figure}[h!]
    \centering
    \includegraphics[width=1.0\linewidth]{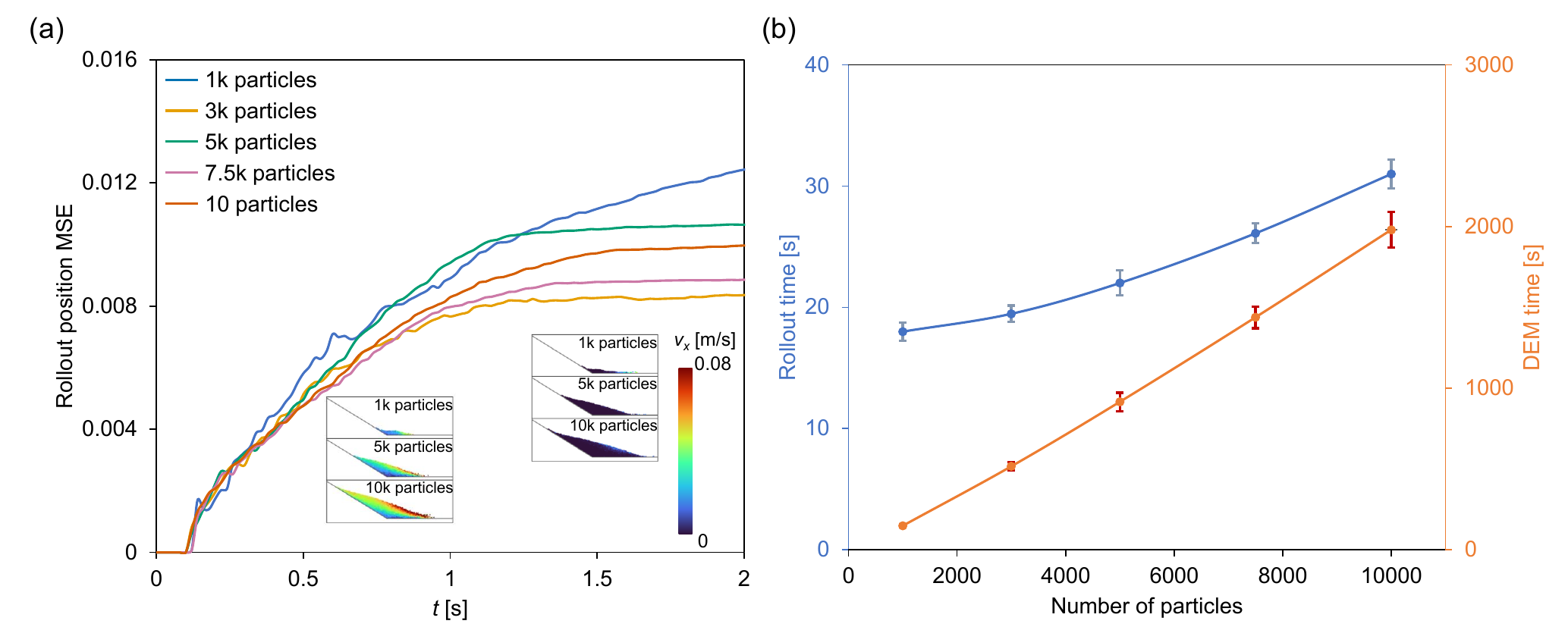}
    \caption{Prediction error and runtime for granular flows with different particle numbers. (a) Evolution of HCG-GNN rollout position error over 2~s. The 7500- and 10000-particle assemblies exceed the maximum training size of 5000 particles. The inset snapshots show representative predicted particle distributions colored by the streamwise velocity \(v_x\). (b) Wall-clock time for HCG-GNN rollout (left axis) and the corresponding DEM simulation (right axis). Error bars indicate one standard deviation across five repeated runs.}
    \label{fig:Scalability}
\end{figure}

Figure~\ref{fig:Scalability}(a) compares rollout position errors over 2~s for systems containing $N=$ 1000, 3000, 5000, 7500, and 10000 particles. Note that the model was trained only on cases with up to 5000 particles, so the last two cases evaluate its generalizability with larger particle systems. The errors grow during the early stages of motion and approach a plateau for most assemblies, while the 1000-particle case continues to accumulate error and has the largest discrepancy at $t=2$~s. At that time, both larger assemblies fall within the error range of the three particle counts represented in training. This suggests that increasing the particle number to twice the training maximum preserved the level of prediction accuracy observed at smaller particle counts.

Figure~\ref{fig:Scalability}(b) compares the HCG-GNN inference time with the corresponding DEM simulation time. As the particle number increases from \(1000\) to \(10000\), the rollout time increases only from 18 to 31~s, while the DEM computational time increases from approximately \(150\) to 2000~s. The ratio of DEM to HCG-GNN runtime consequently increased from $\sim 8$ times for 1000 particles to $\sim 65$ times for 10000 particles under the benchmark conditions. This difference reflects the distinct computational mechanisms of the two approaches: 1) DEM resolves contact dynamics over smaller time increments, while HCG-GNN performs inference every recorded frame interval; 2) the number of nodes at each coarse graph level is bounded by the number of cells in its grid, even as the particle graph grows, allowing a substantial part of the message passing to be performed on fewer nodes and edges. The small error bars in Figure~\ref{fig:Scalability}(b) further indicate that the measured computational times are stable across the evaluated simulations.

\subsection{Ablation studies}

\subsubsection{Skip connections and hierarchy depth}

\begin{figure}[h!]
    \centering
    \includegraphics[width=1.0\linewidth]{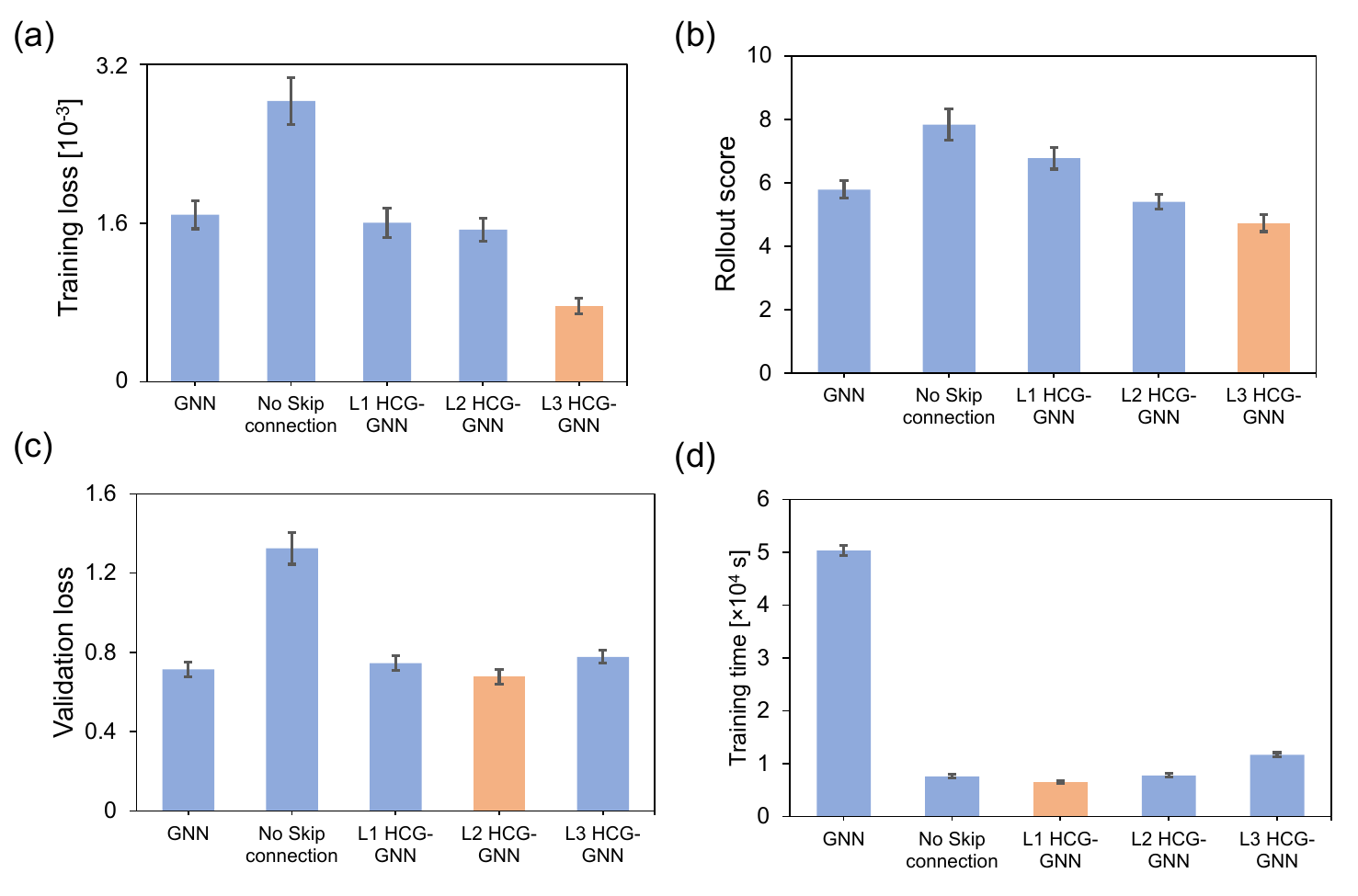}
    \caption{Effects of skip connections and hierarchy depth. The conventional particle-level GNN is compared with a hierarchical variant without skip connections and with HCG-GNN variants containing one, two, or three coarse levels, marked as L1, L2, and L3. Panels present (a) training loss, (b) composite rollout score, (c) validation loss, and (d) total training time. Lower values indicate better performance for all four metrics. The orange bar identifies the best-performing configuration in each panel. Error bars indicate the standard deviation across five repeated runs.}
    \label{fig:module_ablation}
\end{figure}

Figure~\ref{fig:module_ablation} compares the conventional GNN with four hierarchical variants: a model without skip connections and models with one, two, or three coarsening levels, labeled as L1, L2, and L3. The variant without skip connections has the poorest predictive performance, with the highest mean rollout score (7.84) and one-step validation loss (1.33). This deterioration indicates that coarse-scale representations alone are insufficient to recover particle-level dynamics. During coarsening, particles within the same grid cell are aggregated into a shared representation, resulting in the loss of part of the local spatial and interaction information. The skip connections preserve fine-scale particle features and combine them with the upsampled coarse-scale information, which is necessary for maintaining predictive accuracy.

Hierarchy depth affects rollout accuracy and one-step validation loss differently. The one-level model trained fastest (\(6.55\times10^3~\mathrm{s}\)) but had a higher rollout score (\(6.78\)) than the conventional GNN. Adding a second coarse level reduced the score to 5.41 and yielded the lowest one-step validation loss (\(0.678\)) among the variants. A third level further reduced the rollout score to \(4.73\), but raised the validation loss to \(0.778\) and increased training time by $\sim 50\%$ relative to the two-level model. Therefore, we selected the two-level configuration since it improved rollout accuracy over the one-level model while retaining a lower training cost and lower one-step validation loss than the three-level model. Overall, the L2 HCG-GNN provides a favorable balance between predictive performance and computational efficiency and was selected as the primary architecture in this study.

\subsubsection{Physical time integration and loss function}

\begin{figure}[h!]
    \centering
    \includegraphics[width=1.0\linewidth]{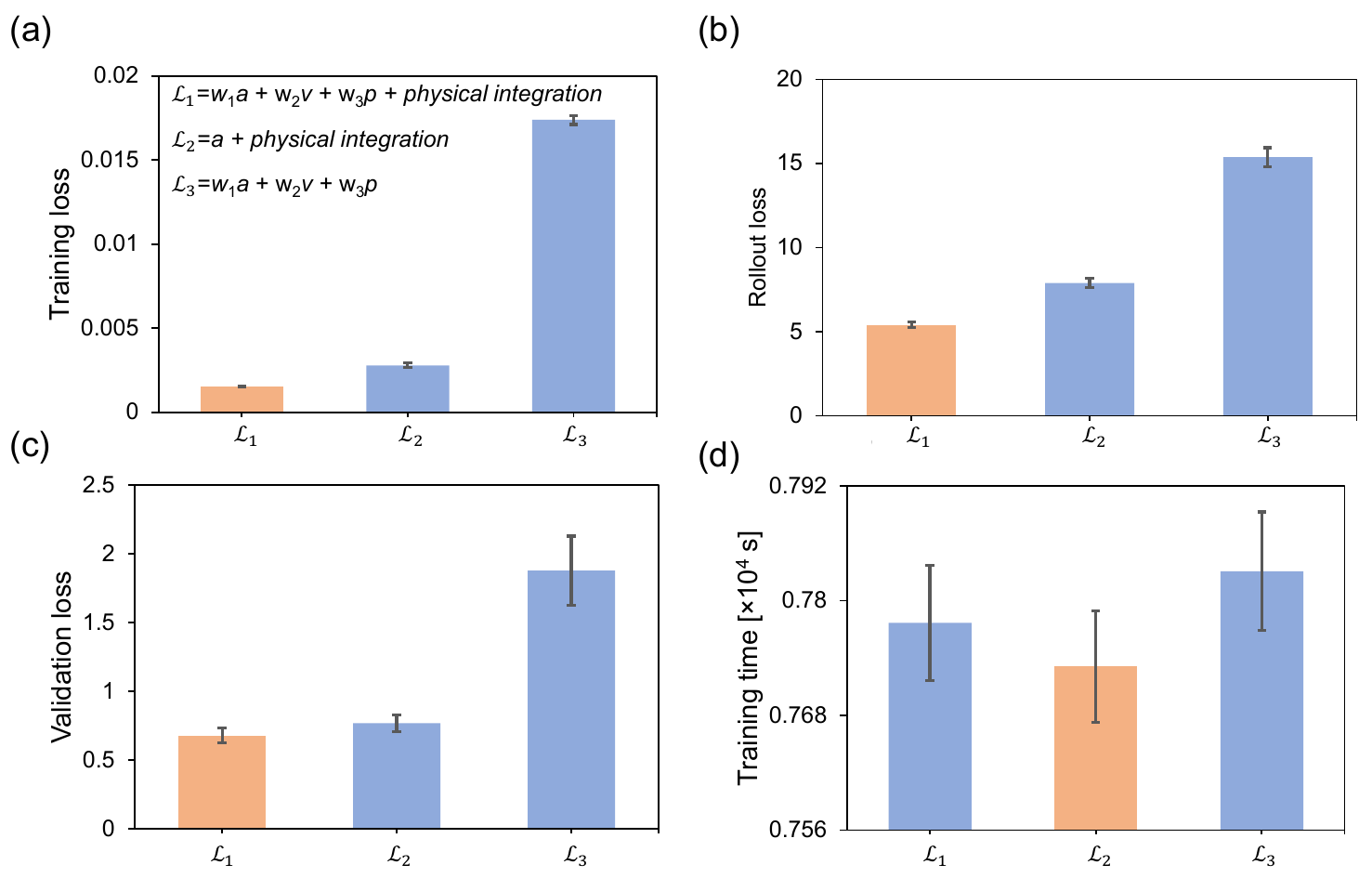}
    \caption{Effects of physical time integration and loss function on rollout performance and training cost. Configurations \(\mathcal{L}_1\) and \(\mathcal{L}_2\) use explicit time integration with the combined and acceleration-only training objectives, respectively. Configuration \(\mathcal{L}_3\) uses the combined objective without explicit integration. Panels present (a) training loss under each configuration's objective, (b) rollout loss, (c) validation loss, and (d) total training time. Lower values indicate better performance for all metrics. The orange bar identifies the best-performing configuration in each panel. Error bars indicate the standard deviation across five repeated runs.}
    \label{fig:input and loss ablation}
\end{figure}

Figure~\ref{fig:input and loss ablation} compares 3 training configurations: 1) the first (\(\mathcal{L}_1\)) uses the time integration scheme in Section~\ref{sec:Physical time integration} and includes acceleration, velocity, and position errors in the training objective; 2) the second (\(\mathcal{L}_2\)) uses the same integration scheme with an acceleration-only objective; 3) the third (\(\mathcal{L}_3\)) retains all three loss terms and omits the explicit integration step. Removing physical integration resulted in the largest performance deterioration: the rollout and validation losses increased from \(5.3\) and \(0.67\) for \(\mathcal{L}_{1}\) to \(15.2\) and \(1.88\) for \(\mathcal{L}_{3}\), respectively. Physical integration imposes a deterministic relationship among acceleration, velocity, and position, thereby reducing the dynamic relationships that must be learned by the network and limiting the accumulation of inconsistent particle states during autoregressive rollout.

Among the two configurations using explicit integration (\(\mathcal{L}_{1}\) and \(\mathcal{L}_{2}\)), including velocity and position errors reduced the rollout metric from \(7.9\) to \(5.3\), a decrease of \(33\%\). This comparison indicates that the weighting of one-step errors influenced the accuracy obtained over successive updates.
Since the three configurations optimize different objectives, their absolute training losses in Figure~\ref{fig:input and loss ablation}(a) are not strictly comparable. Nevertheless, their training times were similar, remaining within \(7.7\text{--}7.8\times10^{3}~\mathrm{s}\), showing that physical integration and the auxiliary loss terms introduce negligible computational overhead. Overall, the combination of physical integration with acceleration, velocity, and position losses provides the best predictive performance among the evaluated configurations.

\subsubsection{Resolution of the coarse graphs}

Table~\ref{tab:grid_ablation_results} compares three grid resolutions for the two-level architecture. Relative to the medium configuration, refining the grids reduced the reported rollout position error from 0.0099 to 0.0094 (5.1\%), but increased training time by 28.6\% and GPU memory usage by 24.5\%. Coarsening the grids reduced training time by 22.8\% and memory usage by 19.3\%, while increasing the position error to 0.0117 (18.2\%). Therefore, the medium configuration captured most of the accuracy obtained with the fine grids at substantially lower cost. We used this configuration, with \(32\times6\times16\) cells at \(L_1\) and \(12\times4\times8\) at \(L_2\), for the principal comparisons.

\begin{table}[h!] 
\centering 
\small 
\setlength{\tabcolsep}{4pt} 
\caption{Effects of coarse-grid resolution on predictive performance and computational cost.} 
\label{tab:grid_ablation_results} 

\begin{tabularx}{\textwidth}{ 
l 
>{\centering\arraybackslash}p{2.0cm} 
>{\centering\arraybackslash}p{2.0cm} 
>{\centering\arraybackslash}p{3.0cm}
>{\centering\arraybackslash}p{2.3cm} 
>{\centering\arraybackslash}p{1.6cm} 
} 
\toprule 
Grid setting 
& L1 grid 
& L2 grid 
& \shortstack{Rollout position\\MSE} 
& \shortstack{Training time\\(500 epochs)} 
& \shortstack{GPU memory\\(GB)} \\ 
\midrule 

Fine 
& \((40,8,20)\) 
& \((16,4,10)\) 
& \(0.0094\) 
& \(2{:}46{:}40\) 
& \(26.957\) \\ 
 
Medium 
& \((32,6,16)\) 
& \((12,4,8)\) 
& \(0.0099\) 
& \(2{:}09{:}37\) 
& \(21.659\) \\ 
 
Coarse 
& \((24,4,12)\) 
& \((8,2,6)\) 
& \(0.0117\) 
& \(1{:}40{:}01\) 
& \(17.475\) \\ 

\bottomrule 
\end{tabularx} 
\end{table}

\section{Conclusions}

This study developed a Hierarchical Coarse-Grained Graph Neural Network (HCG-GNN) for efficient particle-resolved prediction of granular flow dynamics by combining particle-level message passing, hierarchical graph coarsening, multiscale feature transfer, skip connections, and physical time integration. The model exchanges information directly between neighboring grains before pooling their features and again after information from two coarser graph levels returns to the particles. Main findings are summarized as follows:

\begin{itemize}

\item HCG-GNN reproduced the spatiotemporal evolution of granular flow on inclined planes and the rollout captured key flow characteristics, including flow front propagation, lateral spreading, deposition profile, particle redistribution, free surface evolution, internal shear behavior, and energy conversion. The advancing front showed the largest local discrepancies: particle accelerations were smoother than in DEM during the early stage, and the predicted surface elevation exceeded DEM predictions by $\sim 1$ median grain diameter during spreading.

\item Compared with a conventional GNN that performed all 10 message-passing steps on the particle graph, HCG-GNN achieved comparable multistep prediction accuracy with 5 of its 7 steps performed on coarse graphs. It trained 6.47 times faster and used 46.2\% less GPU memory under the tested configuration.

\item Both preserving fine-scale features through skip connections and updating velocity and position from the predicted acceleration improved rollout accuracy. Increasing the hierarchy from 2 to 3 coarse levels further improved multistep predictions but required 50\% more training time. Within the two-level architecture, refining the coarse grids reduced the reported particle-position error by only 5.1\% while increasing training time by 28.6\%. These comparisons support the choice of two coarse levels at the intermediate grid resolution for the flows examined here.

\item The model trained on up to 5000 grains performed well on 7500- and 10000-grain systems. For the 10000-grain case, HCG-GNN required $\sim 31$~s, compared with 2000~s for DEM under the benchmark conditions, corresponding to a 65-fold reduction in simulation time.
\end{itemize}

This study demonstrates that particle-resolved granular flows can be performed when most of the computation that links local motion to collective spreading takes place on coarser graphs. This capability opens a path toward particle-resolved inverse design, in which a prescribed runout limit or discharge rate guides the search for suitable geometries and operating conditions, and grain dynamics reveal the mechanisms behind competing designs.

\section*{CRediT authorship contribution statement}
\textbf{Kuang Shi}: Methodology, Validation, Formal analysis, Investigation, Data curation, Visualization, Writing - original draft.
\textbf{Qingqi Zhao}: Methodology, Software, Validation, Formal analysis, Writing – review \& editing.
\textbf{Teng Man}: Software, Writing - review \& editing. 
\textbf{Sergio Andres Galindo-Torres}: Software, Writing - review \& editing.
\textbf{Yimin Lu}: Conceptualization, Supervision, Project administration, Funding acquisition, Writing – review \& editing.

\section*{Acknowledgements}
The authors acknowledge the Startup Funds from Texas Tech University (TTU) and the high performance computing resources from the Texas Advanced Computing Center (TACC) at The University of Texas at Austin. The DEM simulations were conducted using MechSys, an open-access multiphysics simulation library, and its source code is available at \url{https://github.com/Axtal/mechsys.git}. The authors thank Dr. Krishna Kumar for his guidance, insights, and suggestions on this work.

\bibliography{references}

\newpage

\appendix

\section{Conventional GNN}
\label{app:conventional GNN}

The conventional particle level GNN is used as the baseline model for comparison with the proposed HCG-GNN. It uses the same particle level input features, edge features, acceleration prediction target, and physical time integration scheme as the HCG-GNN. The main difference is that all message passing operations are performed directly on the original particle graph, without graph coarsening, pooling, upsampling, or skip connections.

At time \(t\), the particle system is represented as a graph
\(\mathcal{G}^{t}=(\mathcal{V}^{t},\mathcal{E}^{t})\), where each node represents a particle. Two particles are connected when their separation distance is within the connectivity radius \(R_c\),
\begin{align}
(j,i)\in\mathcal{E}^{t}
\qquad \mathrm{if} \qquad
\left\|
\mathbf{p}_{i}^{t}
-
\mathbf{p}_{j}^{t}
\right\|_2
\leq R_c,
\qquad j\neq i.
\end{align}
The resulting particle node and edge features, denoted by \(\mathbf{x}_{i}^{t}\) and \(\mathbf{e}_{ji}^{t}\), respectively, are encoded into latent representations as
\begin{align}
\mathbf{h}_{i}^{0}
&=
\phi_{v}^{\mathrm{enc}}
\left(
\mathbf{x}_{i}^{t}
\right),
\\
\mathbf{g}_{ji}^{0}
&=
\phi_{e}^{\mathrm{enc}}
\left(
\mathbf{e}_{ji}^{t}
\right),
\end{align}
where \(\phi_{v}^{\mathrm{enc}}\) and \(\phi_{e}^{\mathrm{enc}}\) denote the node and edge encoders.

Information is then propagated over the particle graph through repeated message passing. At message passing step \(m\), the edge feature from sender node \(j\) to receiver node \(i\) is updated, the incoming edge features are aggregated, and the node feature is subsequently updated according to
\begin{align}
\mathbf{g}_{ji}^{m+1}
&=
\mathbf{g}_{ji}^{m}
+
\psi_{e}^{m}
\left(
\left[
\mathbf{h}_{j}^{m},
\mathbf{h}_{i}^{m},
\mathbf{g}_{ji}^{m}
\right]
\right),
\\
\mathbf{m}_{i}^{m+1}
&=
\sum_{j:(j,i)\in\mathcal{E}^{t}}
\mathbf{g}_{ji}^{m+1},
\\
\mathbf{h}_{i}^{m+1}
&=
\mathbf{h}_{i}^{m}
+
\psi_{v}^{m}
\left(
\left[
\mathbf{h}_{i}^{m},
\mathbf{m}_{i}^{m+1}
\right]
\right),
\end{align}
where \(\psi_{e}^{m}\) and \(\psi_{v}^{m}\) are the edge and node update networks, respectively. The residual connections retain the existing latent information while incorporating newly aggregated interaction information. After \(M\) message passing steps, information can propagate through the \(M\)-hop neighborhood of each particle.

Finally, the latent particle representation after the last message passing step is decoded into the normalized three-dimensional particle acceleration,
\begin{align}
\widehat{\mathbf{a}}_{i,n}^{t}
=
\phi_{\mathrm{dec}}
\left(
\mathbf{h}_{i}^{M}
\right),
\end{align}
where \(\phi_{\mathrm{dec}}\) denotes the output decoder. The predicted acceleration is subsequently transformed to the physical scale and integrated to update the particle velocity and position using the same procedure described in Section~\ref{sec:Physical time integration}. In contrast to the proposed HCG-GNN, the conventional GNN relies entirely on repeated message passing over the fine particle graph to enlarge its receptive field.

\section{Additional granular flow prediction results}
\label{app:Additional granular flow prediction results}

\begin{figure}[h!]
    \centering
    \includegraphics[width=1.0\linewidth]{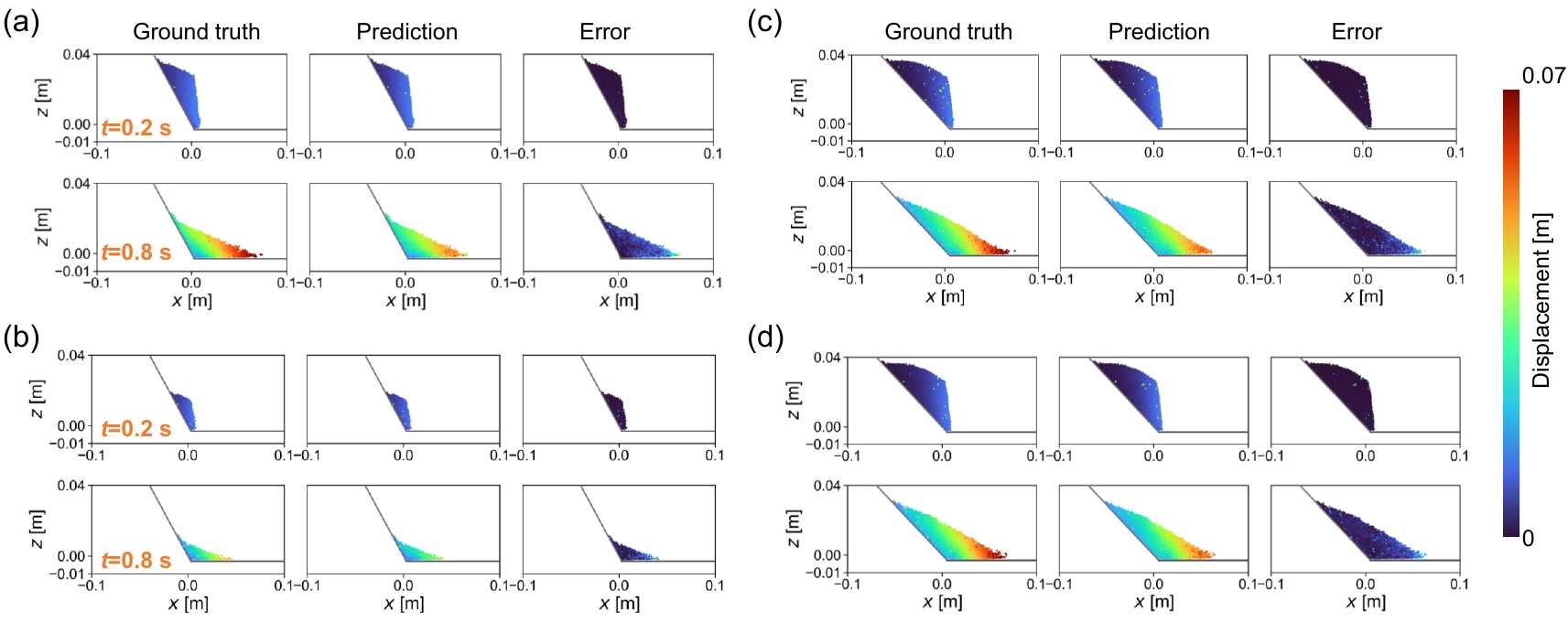}
    \caption{Comparison of particle displacement fields for different granular flow cases. (a) \(\theta=45^\circ\), \(N=3000\), seed \(=5000\); (b) \(\theta=45^\circ\), \(N=1000\), seed \(=5000\); (c) \(\theta=30^\circ\), \(N=5000\), seed \(=5000\); and (d) \(\theta=30^\circ\), \(N=5000\), seed \(=7000\).}
    \label{fig:additional_cases}
\end{figure}

Figure~\ref{fig:additional_cases} presents additional comparisons between the HCG-GNN predictions and DEM results for different inclination angles, particle numbers, and initial conditions. In all four cases, the predicted displacement fields show good agreement with the DEM results at both early and later stages of the flow. The model reproduces the overall runout distance, flow shape, and spatial distribution of particle displacement, while the corresponding errors remain relatively small over most of the granular domain. These results further demonstrate that the proposed HCG-GNN maintains good predictive accuracy across different granular flow conditions.

\begin{figure}
    \centering
    \includegraphics[width=0.6\linewidth]{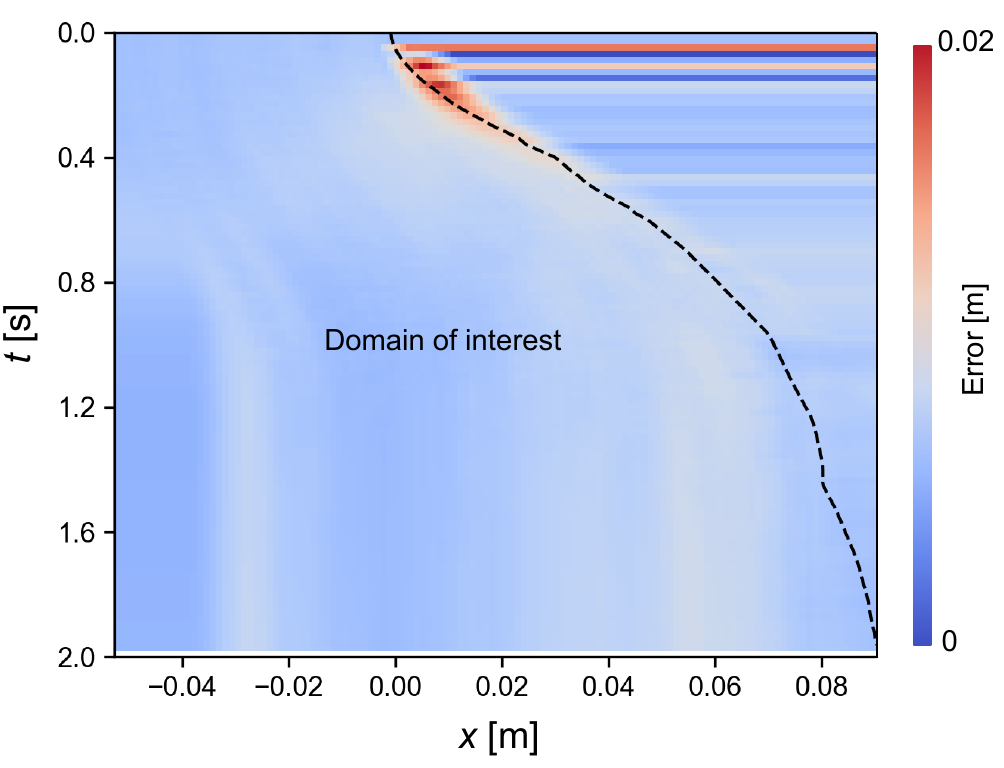}
    \caption{Pile contour error over time}
    \label{fig:pile contour error over time}
\end{figure}

Figure~\ref{fig:pile contour error over time} presents the pile contour error over time. The black dashed line denotes the advancing front of the granular flow, and the region behind this line represents the domain occupied by particles. As the flow propagates downslope and spreads along the horizontal plate, the contour error remains consistently small throughout the simulation. This indicates that the model accurately reproduces the temporal evolution of the overall pile geometry and flow front propagation.

\section{Definitions of physical evaluation metrics}
\label{app:physical_metrics}

Several complementary metrics were used to evaluate the autoregressive predictions beyond visual comparison. These metrics quantify particle level trajectory errors, longitudinal mass redistribution, free surface geometry, internal shear deformation, and energy evolution. Unless otherwise stated, the predicted and DEM particle coordinates were converted to SI units before evaluation, and the same spatial bins and processing parameters were used for both datasets.

\subsection{Longitudinal particle-fraction distribution}

The longitudinal distribution of the granular material is evaluated by dividing the common \(x\)-coordinate range of the DEM and predicted particle clouds into \(N_b\) equally spaced bins. The particle fraction in bin \(k\) is calculated as
\begin{align}
f_k^t
=
\frac{1}{N}
\sum_{i=1}^{N}
\mathbb{I}
\left(
x_k
\leq
x_i^t
<
x_{k+1}
\right).
\end{align}
where \(f_k^t\) denotes the fraction of particles contained in the \(k\)-th longitudinal bin at time \(t\), \(x_k\) and \(x_{k+1}\) are the lower and upper boundaries of the bin, and \(\mathbb{I}(\cdot)\) is the indicator function, which equals one when the stated condition is satisfied and zero otherwise.

In the present evaluation, \(N_b=80\) bins were used, and the same bin boundaries were applied to the DEM and predicted configurations. Because the metric is normalized by the total number of particles, it describes the fraction of the particle population within each longitudinal interval rather than the local solid volume fraction. It therefore permits direct comparison between systems containing different numbers of particles.

\subsection{Upper surface profile}

The upper surface of the granular deposit is extracted from the particle elevations within the same longitudinal bins. To reduce sensitivity to isolated particles above the main granular body, the surface height is defined using the 95th percentile:
\begin{align}
h_k^t
=
Q_{0.95}
\left(
\left\{
z_i^t
\;\middle|\;
x_k
\leq
x_i^t
<
x_{k+1}
\right\}
\right).
\end{align}
where \(h_k^t\) denotes the estimated upper surface elevation in the \(k\)-th bin, \(Q_{0.95}(\cdot)\) denotes the 95th-percentile operator, and \(z_i^t\) is the vertical coordinate of particle \(i\). Bins containing no particles are excluded from the profile comparison.

The difference between the predicted and DEM surface profiles can be summarized using the mean absolute height error:
\begin{align}
E_h^t
=
\frac{1}{|\mathcal{B}^t|}
\sum_{k\in\mathcal{B}^t}
\left|
\widehat{h}_k^t-h_k^t
\right|.
\end{align}
where \(E_h^t\) denotes the surface profile error, \(\widehat{h}_k^t\) and \(h_k^t\) are the predicted and DEM surface heights, respectively, and \(\mathcal{B}^t\) is the set of bins occupied in at least one of the two particle configurations.

\subsection{Horizontal shear strain rate profile}

Particle velocities are reconstructed from the position trajectories. For an internal frame, the velocity is evaluated using a central finite difference:
\begin{align}
\mathbf{v}_i^t
=
\frac{
\mathbf{p}_i^{t+1}
-
\mathbf{p}_i^{t-1}
}{
2\Delta t
}.
\end{align}
where \(\mathbf{v}_i^t\) denotes the velocity of particle \(i\) at time \(t\), and \(\Delta t\) is the physical interval between two recorded frames. Forward and backward differences are used at the first and final frames, respectively.

To obtain a depth-dependent profile, particles within a narrow longitudinal sampling region are divided into \(N_z\) vertical bins. The mean streamwise speed in vertical bin \(k\) is calculated as
\begin{align}
\overline{v}_{x,k}^{\,t}
=
\frac{1}{N_k^t}
\sum_{i\in\mathcal{S}_k^t}
\left|
v_{x,i}^t
\right|.
\end{align}
where \(\overline{v}_{x,k}^{\,t}\) denotes the mean streamwise speed in bin \(k\), \(v_{x,i}^t\) is the \(x\)-direction velocity of particle \(i\), \(\mathcal{S}_k^t\) is the set of particles located in the corresponding longitudinal sampling region and vertical bin, and \(N_k^t=|\mathcal{S}_k^t|\) is the number of particles in that set.

The horizontal shear strain rate magnitude is subsequently estimated from the vertical gradient of the mean streamwise velocity:
\begin{align}
\dot{\gamma}_{xz,k}^{\,t}
=
\left|
\frac{
\partial\overline{v}_x^t
}{
\partial z
}
\right|_{z=z_k}.
\end{align}
where \(\dot{\gamma}_{xz,k}^{\,t}\) denotes the shear strain rate magnitude at the center \(z_k\) of vertical bin \(k\). The derivative is evaluated numerically from the binned velocity profile.

For the profiles presented in the main text, particles satisfying
\(|x-x^\ast|\leq0.0035~\mathrm{m}\) were selected, where \(x^\ast=0.02~\mathrm{m}\). The vertical domain was divided into 48 bins, and a five-point moving average filter was applied to reduce fluctuations caused by the finite number of particles in each bin. Identical binning and smoothing procedures were applied to the DEM and predicted trajectories.

\subsection{Energy evolution}

The mass of each spherical particle is calculated from its radius and material density:
\begin{align}
m_i
=
\rho_p
\frac{4}{3}
\pi r_i^3.
\end{align}
where \(m_i\) denotes the mass of particle \(i\), \(\rho_p\) is the particle density, and \(r_i\) is the particle radius. All quantities must be expressed in consistent SI units.

The gravitational potential and kinetic energies of the granular assembly are calculated as
\begin{align}
E_p^t
&=
\sum_{i=1}^{N}
m_i g
\left(
z_i^t-z_{\mathrm{ref}}
\right),
\\
E_k^t
&=
\frac{1}{2}
\sum_{i=1}^{N}
m_i
\left\|
\mathbf{v}_i^t
\right\|_2^2.
\end{align}
where \(E_p^t\) and \(E_k^t\) denote the gravitational potential and kinetic energies at time \(t\), respectively, \(g\) is the gravitational acceleration, and \(z_{\mathrm{ref}}\) is the reference elevation used consistently for both the DEM and predicted trajectories.

The initial total mechanical energy is defined as
\begin{align}
E_0
=
E_p^0+E_k^0.
\end{align}
where \(E_0\) denotes the initial mechanical energy, and \(E_p^0\) and \(E_k^0\) are the potential and kinetic energies at the initial evaluated frame, respectively.

Because the learned simulator does not explicitly predict contact forces, the dissipated energy is estimated from the mechanical energy balance:
\begin{align}
E_d^t
=
E_0-E_p^t-E_k^t.
\end{align}
where \(E_d^t\) denotes the cumulative energy loss inferred from the reduction in mechanical energy. It represents the combined effects of particle--particle collisions, friction, particle--boundary interactions, and numerical or prediction errors, rather than a direct integration of individual contact dissipation terms.

The normalized energy components are calculated as
\begin{align}
\widetilde{E}_p^t
=
\frac{E_p^t}{E_0},
\qquad
\widetilde{E}_k^t
=
\frac{E_k^t}{E_0},
\qquad
\widetilde{E}_d^t
=
\frac{E_d^t}{E_0}.
\end{align}
where \(\widetilde{E}_p^t\), \(\widetilde{E}_k^t\), and \(\widetilde{E}_d^t\) denote the normalized potential, kinetic, and dissipated energies, respectively. This normalization permits comparison of the energy conversion processes between the DEM and HCG-GNN trajectories independently of the total particle mass.

\newpage

\end{document}